\documentclass[onecolumn,authoryear]{els-mrw}

\usepackage{amsmath,amssymb,amsfonts,amsthm,makeidx,graphicx}
\usepackage{txfonts}
\usepackage{helvet}
\usepackage{wrapfig}
\usepackage{tabularx}
\usepackage{hyperref}

\newcommand*\aap{A\&A}

\newcommand*\aapr{A\&A~Rev.}

\newcommand*\actaa{Acta Astron.}
\newcommand*\aj{AJ}

\newcommand*\apj{ApJ}
\newcommand*\apjl{ApJL}

\newcommand*\apjs{ApJS}

\newcommand*\apss{Ap\&SS}
\newcommand*\araa{ARA\&A}

\newcommand*\mnras{MNRAS}
\newcommand*\na{New A}

\newcommand*\nat{Nature}

\newcommand*\pasp{PASP}
\newcommand*\physrep{Phys.~Rep.}

\newcommand*\prd{Phys.~Rev.~D}

\newcommand*\prl{Phys.~Rev.~Lett.}

\newcommand*\zap{ZAp}

\begin{document}

\chapter{Evolution of binary stars}\label{intsn}

\author[1]{Pablo Marchant}

\address[1]{\orgname{Institute of Astronomy}, \orgdiv{KU Leuven}, \orgaddress{Celestijnenlaan 200D, B-3001 Leuven, Belgium.}}

\articletag{Chapter Article tagline: update of previous edition,, reprint..}

\maketitle

\begin{glossary}[Glossary and nomenclature]
\begin{tabular}{@{}lp{34pc}@{}}
Binary star & Gravitationally bound pair of stars that orbit each other.  \\ 
Close binary star & A binary star with an orbital separation that is sufficiently small to allow for strong interaction between its components.\\
Contact binary & A binary where both components extend beyond their Roche lobes, sharing an outer envelope in equilibrium that closely aligns with a Roche equipotential.\\
Common envelope evolution & Dynamically unstable stage of evolution where the outer layers of a star engulf their binary companion.\\
Detached binary & Binary system where neither component overfills its Roche lobe.\\
Roche lobe & Region in space where co-rotating material is bound to a star in a binary system. Each component of a binary has its own Roche-lobe that can be described as an equipotential of the Roche-potential.\\
Roche-lobe overflow & Name given to the stage where a star extends beyond its own Roche-lobe.\\
Roche potential & An effective potential describing the combined effect of the centrifugal force and the gravitational pull of both components in a binary system. Can also be defined for a single rotating star.\\
Semi-detached binary & Binary system where only one component experiences Roche-lobe overflow and transfers mass to its companion.\\
\end{tabular}
\end{glossary}

\begin{abstract}[Abstract]
Binary stars are pairs of stars that are gravitationally bound, providing in some cases accurate measurements of their masses and radii. As such, they serve as excellent testbeds for the theory of stellar structure and evolution. Moreover, binary stars that orbit each other at a sufficiently small distance will interact during their lifetimes, leading to a multitude of different evolutionary pathways that are not present in single star evolution. Among other outcomes, this can lead to the production of stellar mergers, rejuvenated and chemically contaminated accreting stars, stars stripped of their hydrogen envelopes and gravitational wave sources. For stars massive enough to undergo a supernova, binary interaction is expected to impact the evolution of most of them, making the understanding of binary evolution a critical element to comprehend stellar populations and their impact at large scales.
\end{abstract}

\begin{BoxTypeA}[]{Key points}
\begin{itemize}
\item Binary star evolution leads to complex evolutionary pathways, many of which involve complex 3D hydrodynamical phenomena. Nevertheless, understanding these processes in detail is key to study full stellar populations.
\item Depending on the properties of a binary system, timescales for interaction processes can range between billions of years to hours. Timescales at evolved stages can become even shorter.
\item As a consequence of binary interaction, stars can become stripped of their outer layers, acquire large amounts of mass, or merge with their companions. The resulting binary products are not always easily identifiable.
\item Various new observational constraints to binary evolution theories are becoming available, in large part due to extensive surveys with clearly understood biases. One particularly rapid growing area is gravitational wave astrophysics, with current observations probing the merging remnants of massive stars.
\end{itemize}
\end{BoxTypeA}

\section{Introduction}\label{intro}
Stars are the building blocks of galaxies, driving the chemical evolution of their hosts \citep{MaiolinoMannucci2019} and providing feedback at large scales both through radiation and kinetic feedback by stellar winds and supernovae \citep{Hopkins+2012,Agertz+2013}. Particular types of supernovae that arise from binary interactions have also been used as standard candles, playing a key role in cosmology \citep{Riess+1998}. As such, the formation of stars and their evolution is critical to our understanding of the universe at large scales. In many cases, stars are part of binary systems, leading to diverse evolutionary pathways for both low and high mass stars \citep{Tauris2023, Marchant2024}. As it has become clearer that binary interactions are frequent, and not just the source of "stellar exotica", binary evolution now plays a central role in the study of stellar astrophysics.

\begin{figure*}
\begin{center}
\includegraphics[width=\textwidth]{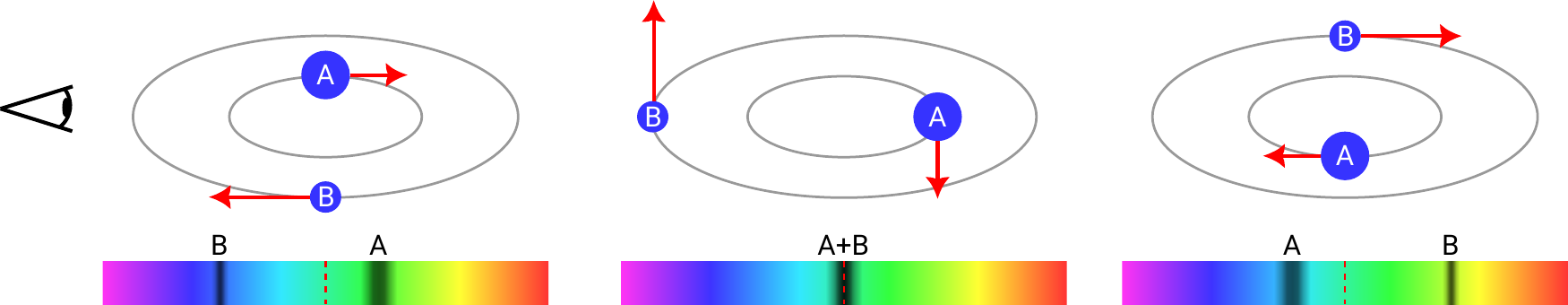} 
\end{center}
    \caption{Illustration of the detection method of binary systems through the measurement of radial velocity variations in spectra. As two stars orbit each other they move in anti-phase, leading to opposite Doppler shifts in spectral lines. The least massive star in a binary system (star B in the illustration) experiences the largest shift in radial velocity through an orbital cycle, and measuring the amplitude of radial velocity variation of both components allows for the inference of the mass ratio. Systems for which spectral lines of both components are detected are referred to as SB2, while systems with only one visible set of spectral lines are known as SB1. Source files to reproduce all figures of this document are available at \url{https://doi.org/10.5281/zenodo.14710862}}
    \label{fig:vr}
\end{figure*}

% Binary detection
There are various complimentary techniques that can be used to identify binaries which are sensitive to different types of systems \citep{Sana2017}. Visual binaries, where the motion of a star as it orbits a companion can be directly resolved, are usually restricted to long orbital periods (see, for instance, the detections of the Gaia mission, \citealt{DR3binaries}). In some cases, interferometric observations even allow to resolve the individual components of interacting binaries, providing a clear view into the structure of these systems \citep{Baron+2012}. Photometry can capture brightness variations associated to the tidal deformation of close binaries, as well as eclipses for binaries that are observed near edge-on (see, for example, the large catalog provided by \citealt{Soszynksi+2016}). Using spectroscopy, radial velocity variations can be inferred from the displacement of spectral lines (see Figure \ref{fig:vr}), which has been used to infer that the majority of massive stars undergo binary interactions \citep{Sana+2012}. Beyond electromagnetic observations, in the past decade the direct detection of gravitational waves using ground-based interferometers has allowed for the identification of merging pairs of black holes and neutron stars \citep{GWTC3}, providing a new avenue to study binary evolution through a glimpse at its last stages. The number of observed gravitational wave sources is expected to increase steeply as detectors are improved \citep{Baibhav+2019}, while the launch of the LISA mission in the coming decade will provide detections in the frequency range relevant for binary white dwarfs \citep{Colpi+2024}. The evolutionary pathways and interaction processes that produce the large variety of binary products observed is an area of very active research.

% Progress, sims
There are still large uncertainties in our understanding of binary star evolution, associated both to uncertain physical processes in single star evolution as well as the complex nature of binary interactions (see \citealt{Marchant2024} and \citealt{Chen+2024} for recent reviews). 
As the equations of stellar structure and evolution are highly non-linear, theoretical stellar evolution is strongly reliant on computational simulations. Detailed models of binary interaction were initially produced just shortly after computers were first used to model single star evolution \citep{KippenhahnWeigert1967}, and present day methods still strongly resemble those first pioneering efforts. As the life of a star involves multiple evolutionary timescales, ranging from minutes to billions of years, one-dimensional approximations are required except for short-lived stages that can be resolved in multi-D hydrodynamical simulations. Strong binary interactions, however, break most symmetries present in a system and often hinder our capacity to produce predictive models. In this document, I provide a brief overview of the basic physics of binary interactions, highlighting the signatures of interaction products as well as some of the diverse pathways that cannot be produced by stars in isolation.

\section{Binary interaction processes}
% Discussion of Roche lobe figure, equation for Roche potential
Binary systems can be broadly classified in terms of their sizes relative to their orbits. A particularly useful tool for this in a circular system is the Roche potential $\Psi$, which is an effective potential in a frame co-rotating with the binary that includes the gravitational field of both stars and the centrifugal force. For the case of hydrostatic equilibrium in the co-rotating frame, the effective gravity $g_\mathrm{eff}=-\nabla\Psi$ must be balanced out by the pressure gradient,
\begin{equation}
    \rho\nabla \Psi = -\nabla P,
\end{equation}
implying that equipotential surfaces must correspond to isobars\footnote{In practice small deviations are expected as a rotating star cannot be in both hydrostatic and thermal equilibrium (see \citealt{vonZeipel1924, Fabry+22}).}, and the photospheres of binary stars are expected to closely follow these surfaces. As most of the mass of a star is centrally concentrated, the Roche potential is often approximated by using the potential of two point masses for each star. For a binary system with masses $m_1$, $m_2$ and separation $a$, the Roche potential can then be written in dimensionless form as
\begin{eqnarray}
\Psi'(x',y',z') \equiv \frac{\Psi a}{Gm_1} = -\frac{1}{r_1'}-\frac{q}{r_2'}-\frac{q+1}{2}\left(x'^2 + y'^2\right), \quad q\equiv\frac{m_2}{m_1},\quad r_1'^2=\left(x'+\frac{q}{1+q}\right)+y'^2+z'^2,\quad r_2'^2=\left(x'-\frac{1}{1+q}\right)+y'^2+z'^2.
\end{eqnarray}
Here $x'$, $y'$ and $z'$ correspond to Cartesian coordinates with their origin at the center of mass and normalized by the orbital separation $a$, with $x'$ oriented along the line that joins both stars and $z'$ oriented perpendicular to the orbital plane. The geometry of the Roche potential is then completely given by the mass ratio of the system.

\begin{figure*}
\begin{center}
\includegraphics[width=\textwidth]{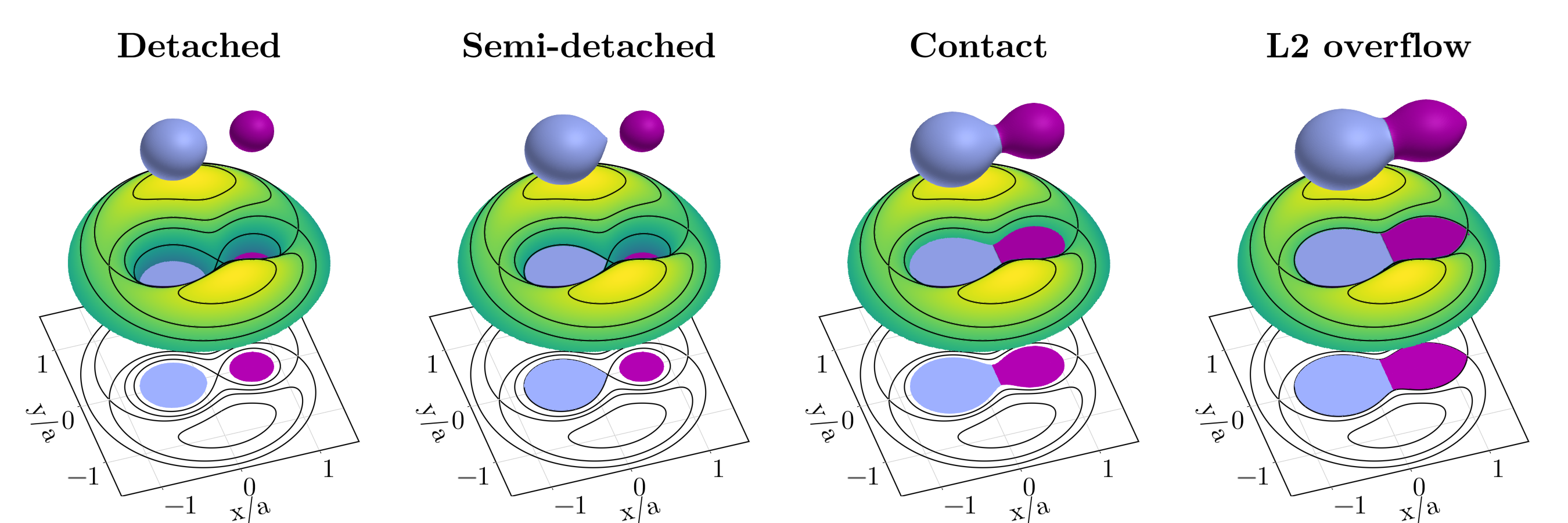} 
\end{center}
    \caption{Different types of binaries with a mass ratio $q=0.5$ in terms of their filling of Roche-equipotential surfaces. Bottom plot shows iso-contours of the Roche potential in the orbital plane, while the 3d-surface indicates the value of the potential in the orbital plane as height in arbitrary units. Light blue and purple surfaces show the three dimensional shape of each star.}\label{fig:roche}
\end{figure*}

Figure \ref{fig:roche} illustrates the shape of the Roche potential for a binary with a mass ratio $q=0.5$. Along the line that joins both components there is a local minimum of $\Psi$ which defines the first Lagrangian point $L_1$. The equipotential surface crossing $L_1$ defines in turn two volumes connected at $L_1$, known as the Roche lobes of each component. As a star grows it can potentially fill its own Roche lobe, at which point material can begin to 'spill' towards its companion, a process known as Roche-lobe overflow. Systems where neither star has filled their Roche lobe are known as detached binaries, while those where one component fills it are known as semi-detached binaries. Stable configurations are also possible when both stars are filling their Roche lobes, in which case their joint surface is expected to follow a single equipotential surface. An arbitrary amount of contact is not possible though, as eventually the equipotential surface corresponding to the second Lagrangian point $L_2$ is filled and material begins to 'spill' outwards, leading to significant angular momentum losses and a potential merger \citep{Kuiper1941, LubowShu1975}. The volume of each Roche lobe goes as $a^3$, scaled by a function of the mass ratio. Often however we consider the volume equivalent radius $R_\mathrm{RL}$, which corresponds to the radius of a sphere with volume equal to that of the Roche lobe. For the case of $m_1$ this is given by
\begin{eqnarray}
    R_\mathrm{RL,1}=f(q)a,\quad f(q)\simeq\frac{0.49q^{2/3}}{0.6q^{2/3}+\ln(q+q^{1/3})},
\end{eqnarray}
where the approximation to $f(q)$ is given by the numerical fit computed by \citet{Eggleton83}. Even before a star in a binary system fills its Roche lobe, interactions can happen, as each component is tidally deformed and can transfer angular momentum to and from the orbit if their rotation is not synchronized. The following subsections describe how stars interact both before and after they fill their Roche lobes.

% timescales of binary interaction

% modelling challenges
\subsection{Tidal interaction}
%overall discussion on tides, mention heartbeats
As highlighted in Figure \ref{fig:roche}, stars that approach their Roche lobe radii can become significantly deformed and deviate from spherical symmetry. This deformation leads to variations in the effective temperature and flux across the stellar surface, resulting in so-called ellipsoidal variability which can be probed with photometry to infer various binary properties \citep{WilsonDevinney1971, PrsaZwitter2005}. In eccentric systems, the particular lightcurve that results has been dubbed as a `heartbeat` \citep{Thompson2012}. Other than photometric variability, tides also play an important role in the transfer of angular momentum between each binary component and the orbital angular momentum itself. Both in eccentric systems and in cases where the rotation of a star does not match its orbital period the tidal bulge induced on a star will not be perfectly aligned along the line connecting both components, which allows for a gravitational torque to be applied in a process known as the equilibrium tide \citep{Darwin1879tides}. The displacement of the tidal bulge is enhanced for lower mass stars, as convection provides a higher effective viscosity and dissipation rate \citep{Zahn1977}. For stars with radiative envelopes, a different effect known as the dynamical tide is thought to be dominant, where oscillation modes that are induced in the tidally distorted star and the particular way in which they dissipate allows for angular momentum transfer \citep{Zahn1975}. In binary evolution models these two effects are often approximated with simple prescriptions (e.g. \citealt{Hurley+2002}), but detailed studies have highlighted the importance and complexity of a proper treatment of tidal effects (e.g. \citealt{FullerLai2012, Sun+2023}).

%circularisation period, dig some of Bob Mathieus work perhaps?

%Darwin instability as example
In some circumstances, tidal interactions can be a dominant process that determines the qualitative evolution of a binary system. One such situation operates in low-mass stars with strong magnetic fields where a stellar wind can be coupled to large distances leading to efficient stellar spin-down. In short period binaries where tides couple the stellar spin to the orbit this magnetic braking leads to efficient removal of angular momentum and a shrinkage of the orbit (\citealt{Huang1966}, see \citealt{VanIvanova2021} for recent work). Another example of tidal interactions that significantly alter stellar evolution is the Darwin instability \citep{Darwin1879}. If we consider a circular binary system where both stars are synchronized to their orbital frequency, the angular momentum contained in the orbit and the spins is given by
\begin{equation}
    J_\mathrm{orb}=m_1m_2\sqrt{\frac{Ga}{m_1+m_2}},\qquad J_\mathrm{spin}=(I_1+I_2)\Omega_\mathrm{orb}=(I_1+I_2)\sqrt{\frac{G(m_1+m_2)}{a^3}},\label{equ:jorb}
\end{equation}
such that the total angular momentum of a synchronized binary, including its orbital and spin components has two terms with a different dependence on the orbital separation,
\begin{equation}
    J_\mathrm{synch}=Aa^{1/2} + Ba^{-3/2},\qquad A\equiv m_1m_2\sqrt{\frac{G}{m_1+m_2}},\qquad B\equiv (I_1+I_2)\sqrt{G(m_1+m_2)}.
\end{equation}
Owing to the different dependence on $a$ of the spin and orbital angular momentum, there is a global minimum $J_\mathrm{min}$ which occurs at an orbital separation
\begin{equation}
    a_\mathrm{min}=\sqrt{3(I_1+I_2)/\mu},
\end{equation}
where $\mu=m_1m_2/(m_1+m_2)$ is the reduced mass of the binary. At this separation $J_\mathrm{orb}/J_\mathrm{spin}=3$, such that a significant fraction of the total angular momentum of the system is contained in the stellar spins.

\begin{figure*}
\begin{center}
\includegraphics[width=\textwidth]{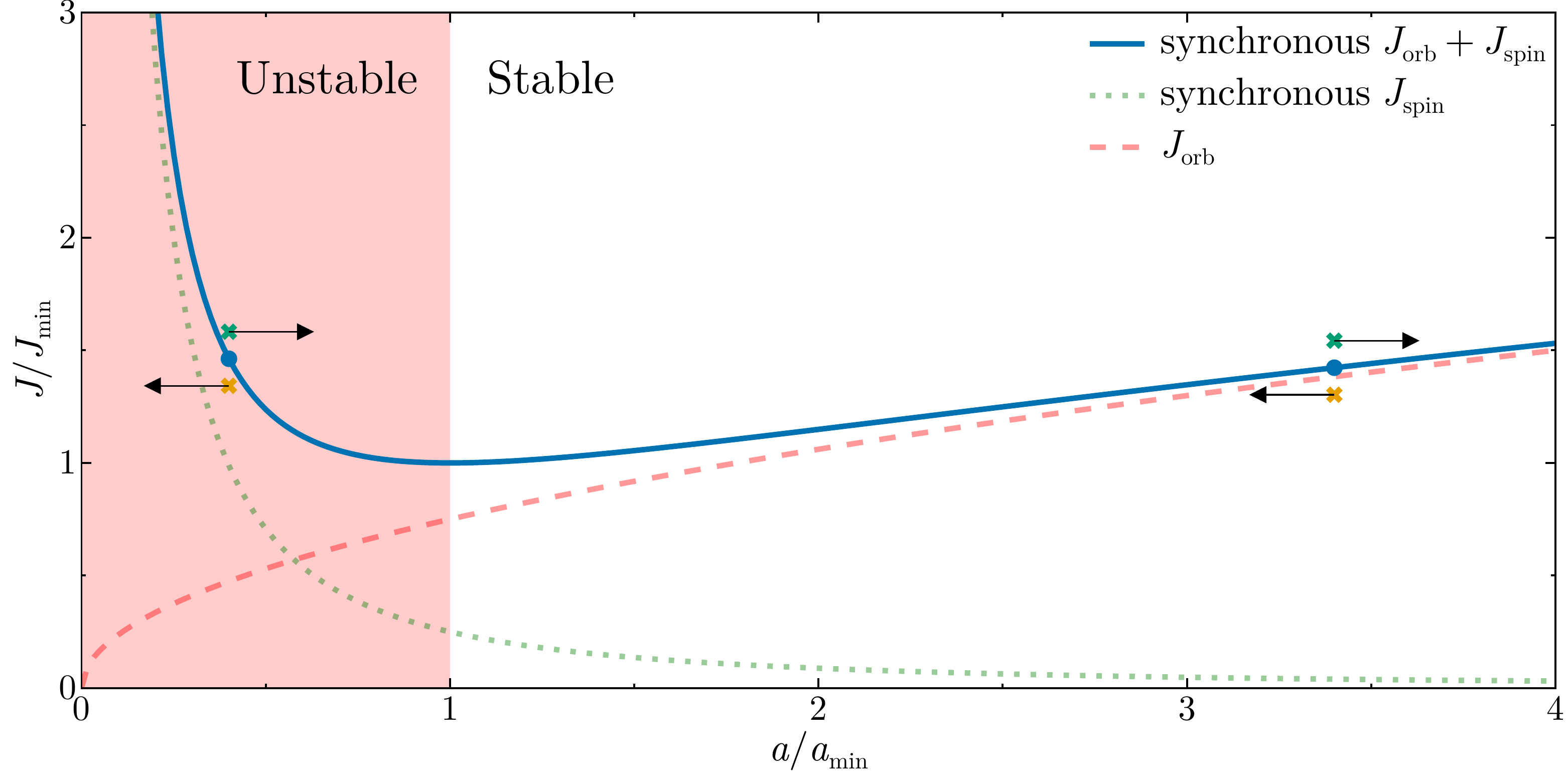} 
\end{center}
    \caption{Mechanism responsible for the Darwin instability. Figure shows the total angular momentum in a synchronized binary system as a function of separation. Angular momentum and separation are normalized by the minimum angular momentum possible in a synchronized binary, $J_\mathrm{min}$, and the corresponding separation where this happens, $a_\mathrm{min}$. Crosses above the blue line correspond to oversynchronous systems, and tidal synchronization makes them evolve to larger separations while conserving the total angular momentum. Crosses below the blue line illustrate undersynchronous systems, which evolve to shorter separations due to tidal synchronization. Non-synchronized systems below $a<a_\mathrm{min}$ cannot evolve to a nearby synchronized configuration, making them unstable.}
    \label{fig:darwin}
\end{figure*}

The Darwin instability arises when we consider systems that are not synchronized, and how their evolution proceeds if $a<a_\mathrm{min}$. An oversynchronous system will have $J/J_\mathrm{synch}>1$ and feed angular momentum to the orbit, resulting in an increase in the orbital separation, while the inverse is true for an undersynchronous system. This is illustrated in Figure \ref{fig:darwin}, where it is shown that systems that are slightly away of synchronicity can adjust and reach a nearby synchronous state if $a>a_\mathrm{min}$. However, if $a<a_\mathrm{min}$ a runaway situation ensues, and particularly for an undersynchronous system this can lead to a merger. This instability can be triggered in an evolving binary system either due to processes that reduce the orbital separation or, as each star evolves, increase their moment of inertia. 
Such runaway process was potentially the cause of the stellar merger V1309 Sco, a transient event whose progenitor was identified as a rapidly inspiraling binary from archival data \citep{Tylenda+2011}.

\subsection{Orbital evolution due to mass transfer}
When a binary reaches a semi-detached stage it will begin mass transfer, and the timescale in which this process occurs will correspond to one of the natural timescales of the donor star. These are the dynamical, thermal and nuclear timescales. During most of the life of a star, these timescales are separated by several orders of magnitude. For massive stars ($M>8M_\odot$) during core-hydrogen burning, their dynamical timescale is in the order of hours, while their thermal timescales are larger than a millennia. Their nuclear timescale on the other hand exceeds a million years, while that of solar-type stars is in the billions. As such, interactions can last for very short periods of time that we can probe with observations, or extend well beyond human history.

Although mass transfer is a complex 3D hydrodynamical process, its impact on orbital evolution can be assesed with some simple assumptions. For the case of a circular orbit, taking the time derivative of the orbital angular momentum from Equation \ref{equ:jorb} one obtains that
\begin{eqnarray}
    \frac{\dot{J}_\mathrm{orb}}{J_\mathrm{orb}} = \frac{\dot{m}_1}{m_1} + \frac{\dot{m}_2}{m_2}-\frac{1}{2}\frac{\dot{m}_1+\dot{m}_2}{m_1+m_2}+\frac{1}{2}\frac{\dot{a}}{a}.\label{equ:jdot}
\end{eqnarray}
In a case where mass is conserved, the orbital angular momentum can still change as it can be transferred between the binary components. Often however the angular momentum contained in the stellar spins is small compared to the orbital angular momentum (opposite to the extreme case of the Darwin instability). In this case one can approximate $\dot{J}_\mathrm{orb}=0$ for conservative mass transfer, or use specific models for how angular momentum is lost as part of the transferred mass is ejected rather than accreted. One example is the assumption of isotropic re-emission, where ejected matter is assumed to remove a specific orbital angular momentum equivalent to that of the accretor. These approximations allow for analytical solutions to Equation \ref{equ:jdot} relating the orbital separation to the mass ratio. In particular if we consider the cases of fully conservative and fully non-conservative mass transfer under isotropic re-emission, it is found that (see \citealt{Soberman+1997,Tauris2023} for a more general set of solutions)
\begin{wrapfigure}{R}{0.5\textwidth}
  \vspace{-0.8cm}
\begin{center}
\includegraphics[width=0.49\textwidth]{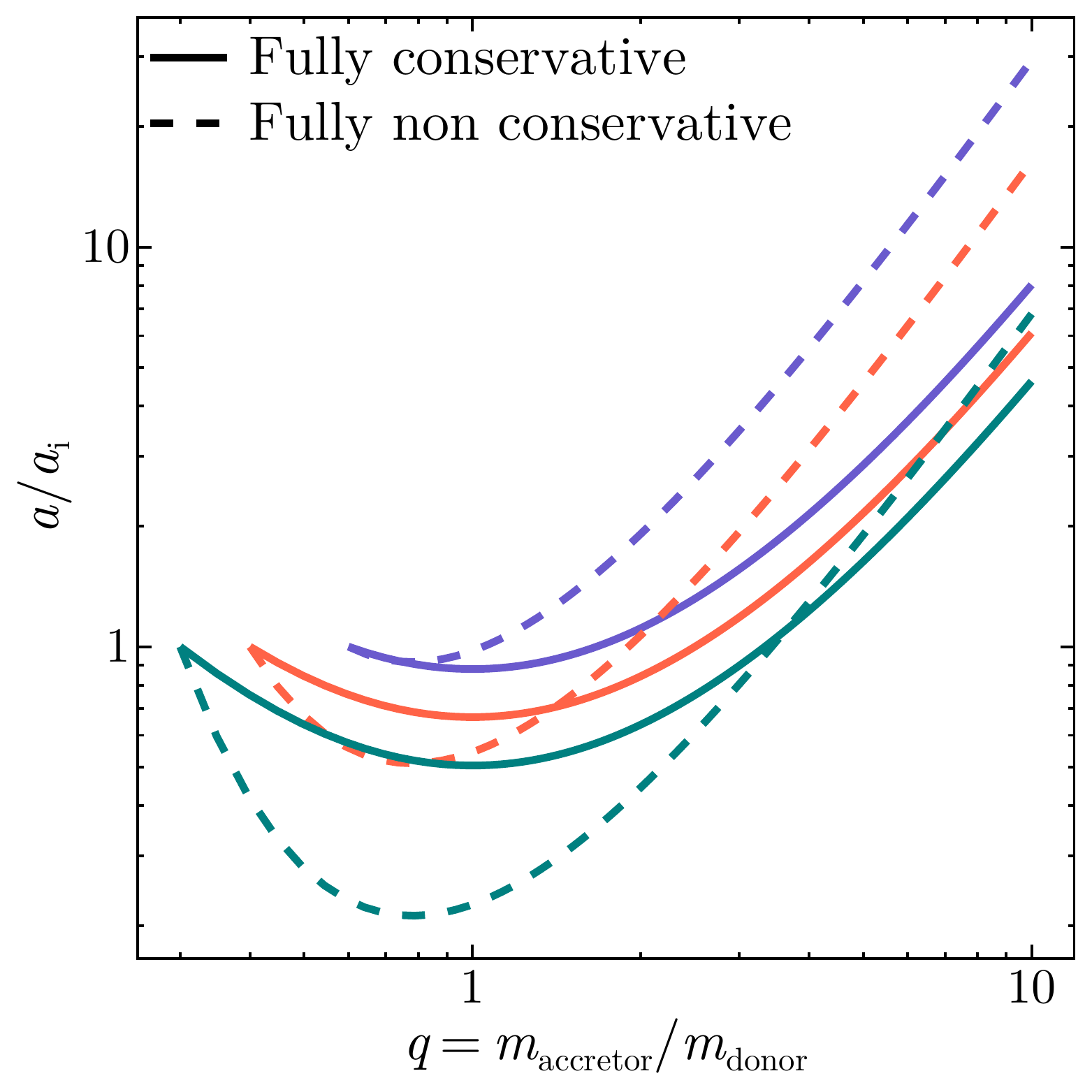} 
\end{center}
\vspace{-0.5cm}
    \caption{Evolution of the orbital separation as a function of mass ratio in a circular binary undergoing mass transfer according to Equation \ref{equ:orbevo}. Solid lines indicate fully conservative mass transfer while dashed lines correspond to fully non-conservative mass transfer under the isotropic re-emission assumption. Different colors correspond to different initial mass ratios $q_\mathrm{i}=0.6$, $0.4$ and $0.3$ at the beginning of the mass transfer stage. The end point at $q=10$ is arbitrary, with the precise point where mass transfer finishes being determined by the internal structure of the donor.}
    \label{fig:accretion}
\end{wrapfigure}
\begin{eqnarray}
\left(\frac{a}{a_\mathrm{i}}\right)_{\mathrm{conservative}} = \left(\frac{q}{q_\mathrm{i}}\right)^{-2}\left(\frac{q+1}{q_\mathrm{i}+1}\right)^{4},\qquad \left(\frac{a}{a_i}\right)_{\mathrm{non\;conservative}}= \left(\frac{q}{q_\mathrm{i}}\right)^3\left(\frac{q+1}{q_\mathrm{i}+1}\right)^{-1}\exp\left(-2[q-q_\mathrm{i}]/[qq_\mathrm{i}]\right),\label{equ:orbevo}
\end{eqnarray}
where $q=m_\mathrm{accretor}/m_\mathrm{donor}$ and $q_\mathrm{i}$ and $a_\mathrm{i}$ are the mass ratio and separation at the onset of interaction. Under most circumstances in a binary system the first stage of interaction will be initiated by the most massive star, as it evolves on a shorter timescale and fills its own Roche lobe, such that $q_\mathrm{i}<1$ (for an exception, see \citealt{deMink+2009, Marchant+2017}).

The resulting evolution of the orbital separation is shown in Figure \ref{fig:accretion}. Conservative systems with $q_\mathrm{i}<1$ shrink as a consequence of mass transfer, reaching a minimum orbital separation when $q=1$. Fully non-conservative evolution behaves similarly, but can also result in significantly more orbital shrinkage during the mass transfer stage and the minimum in orbital separation corresponds to $q=0.781$. Figure \ref{fig:accretion} however does not indicate the timescales at which mass transfer operates. As binaries that start mass transfer with mass ratios further away from unity experience larger orbital shrinkage, this impacts the stability of the mass transfer phase (see \citealt{Soberman+1997} for an overview). Conditions where the donor star can fit in its Roche lobe while retaining thermal equilibrium (where nuclear energy production exactly balances the outgoing luminosity) evolve on the nuclear timescale of the star and are thus more frequent observationally. If thermal equilibrium cannot be sustained, but the stellar interior can still adjust to a hydrostatic configuration that matches the Roche-lobe size, the phase proceeds on the thermal timescale of the star which in most circumstances is orders of magnitude lower than its nuclear timescale. If, even considering an adiabatic response to the mass loss process, the donor cannot adjust to the change in its Roche-lobe size, the process becomes dynamically unstable, leading to a runaway increase in the amount of Roche-lobe overflow with the envelope of the donor engulfing both components. This unstable evolutionary stage is known as common envelope evolution \citep{Paczynski1976}, and is one of the highest sources of uncertainty in binary evolution (see \citealt{Ivanova+2013} and \citealt{RopkeDeMarco2023} for recent reviews).

Mass transfer stages are usually catalogued based on the evolutionary state of the donor star. Based on the notation defined by \citet{KippenhahnWeigert1967} if mass transfer begins while the donor is still a core hydrogen burning star the phase is referred to as Case A mass transfer, while mass transfer after core hydrogen depletion but before core helium depletion is known as Case B. An additional category of Case C mass transfer was later added by \citet{Lauterborn1970} to refer to mass transfer after core helium depletion. A typical scenario for stars that begin interaction in the main sequence is initially a phase of thermally unstable mass transfer, referred to as fast Case A, followed by a thermally stable mass transfer phase known as slow Case A after mass ratio inversion. The prototypical system undergoing slow Case A evolution is the binary system Algol, for which interferometric observations have resolved the binary system and even an outer third companion \citep{Baron+2012}, and as such this phase is usually referred to as the Algol phase. After the depletion of core hydrogen, the remaining hydrogen rich layers expand as a consequence of the contraction of the helium core and shell hydrogen burning leading to another mass transfer phase (referred to as Case AB to denote the previous interaction stage) that strips most of its hydrogen rich layers.

\subsection{Impact of interactions in donors and accretors}
\begin{wrapfigure}{L}{0.51\textwidth}
  %\vspace{-0.8cm}
\begin{center}
\includegraphics[width=0.39\textwidth]{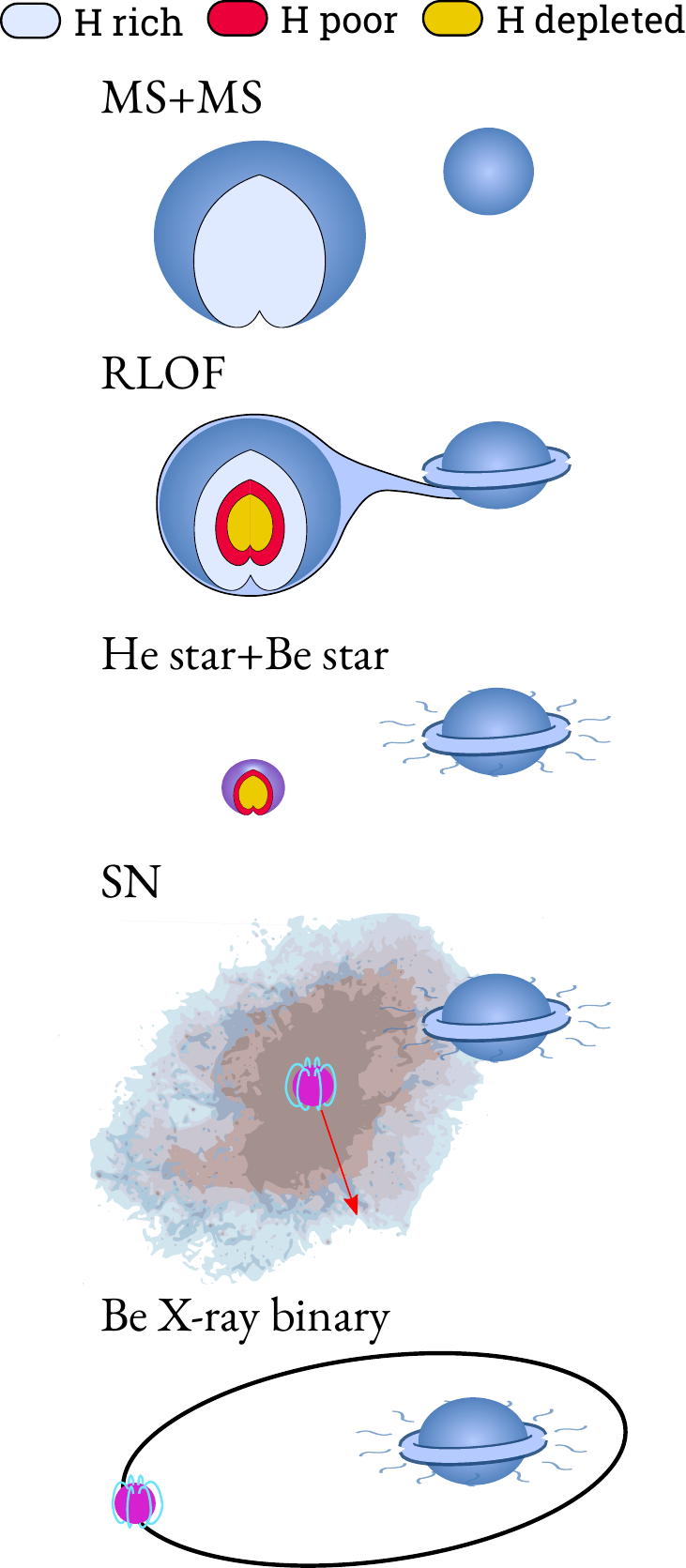} 
\end{center}
%\vspace{-0.5cm}
    \caption{Sequence of evolutionary stages in an interacting massive binary. See text for details. Acronyms: main sequence (MS, core hydrogen burning phase), roche-lobe overflow (RLOF), helium star (He star), supernovae (SN).}
    \label{fig:vdh}
\end{wrapfigure}

% donors, stripped stars, formation of WRs, WDs
After mass transfer completes, the resulting stars (commonly referred to as binary products) are significantly affected by their previous interaction. The different steps that an interacting binary system undergoes are commonly shown using illustrations of intermediate phases as in Figure \ref{fig:vdh}. The evolution shown in Figure \ref{fig:vdh} is a typical case of interaction of a pair of massive binaries, evolving all the way to the point where the initially more massive star undergoes a supernova and forms a neutron star. However, there is an extremely rich variety of outcomes and branching points from this sequence of events, as is shown in \citet{Marchant2024} and \citet{Chen+2024} (with the latter including a very comprehensive set of evolutionary pathways including both low and high mass binary evolution).

The evolutionary sequence in Figure \ref{fig:vdh} shows how the outermost layers of the donor star are removed through mass transfer, resulting in a stripped envelope star that is mostly composed of helium. Depending on the final mass of the stripped star, the resulting object can be a Wolf-Rayet star with strong stellar wind features in its spectrum \citep{Paczynski1967b, Vanbeveren1998, Pauli+2022}, a subdwarf O or B star which burns helium in its core \citep{Han+2002,Pelisoli2020}, all the way down to the so-called extremely low mass white dwarfs that can only be a product of binary interactions \citep{Althaus+2001, Istrate+2016, Li+2019}. Mass transfer is expected to remove most of the hydrogen envelope of the donor, transferring and exposing deep layers of material that have undergone nuclear burning and thus carry important nucleosynthetic signatures. In addition to a depletion of hydrogen in favor of helium, an overabundance of nitrogen together with a depletion of carbon (and potentially oxygen) is a tell-tale sign of material processed through the CNO cycle (e.g. \citealt{Maeder+2014}).

Stripped envelope stars also reach effective temperatures significantly higher than stars during core-hydrogen burning (the main-sequence), leading to a larger fraction of their flux corresponding to ionizing radiation. This can lead to a significant contribution from stripped stars to the integrated UV flux of galaxies as well as contributing to cosmic reionization (e.g. \citealt{Han+2007UV, Stanway+2016, Gotberg+2020}). Despite the expected big role that stripped stars should play in the evolution of the universe, until recently there has been an important gap in observations with a lack of objects observed within the mass range of $2M_\odot-8M_\odot$. Using UV photometry \citet{Drout+2023} were able to unveil a population of such stars in the Magellanic clouds, and future missions such as the UltraViolet EXplorer (UVEX, \citealt{Kulkarni+2021}) are expected to significantly increase the number of detected sources.

% Accretor, mass gain, rejuvanation, blue stragglers
The mass accretor in a binary system also experiences significant changes as a consequence of gaining mass. One particular example is the semidetached binary Algol, where the more massive component is the accretor and it is at an earlier evolutionary stage than its lower mass companion contrary to the expectations of stellar evolution theory. This issue was dubbed the "Algol paradox", and its solution came by noting that the accretor must have been the initially less massive star, made to look younger by the accretion of a significant amount of mass \citep{Crawford1955}. This process of rejuvenation acts two-fold; accretion of mass can lead to the injection of hydrogen rich layers towards the stellar core, but even without that effect, it leads to a star that looks younger for its mass than stars that were initially born as massive at the same time. Some of these rejuvenated stars are easily identifiable within stellar clusters, as they extend beyond the main-sequence turnoff that is set by the most massive stars born single that have evolved beyond core hydrogen burning. As they are seen at higher luminosities and temperatures than the turnoff, these objects are referred to as blue-stragglers \citep{BurbidgeSandage1958}.

\begin{wrapfigure}{R}{0.5\textwidth}
  \vspace{-0.8cm}
\begin{center}
\includegraphics[width=0.49\textwidth]{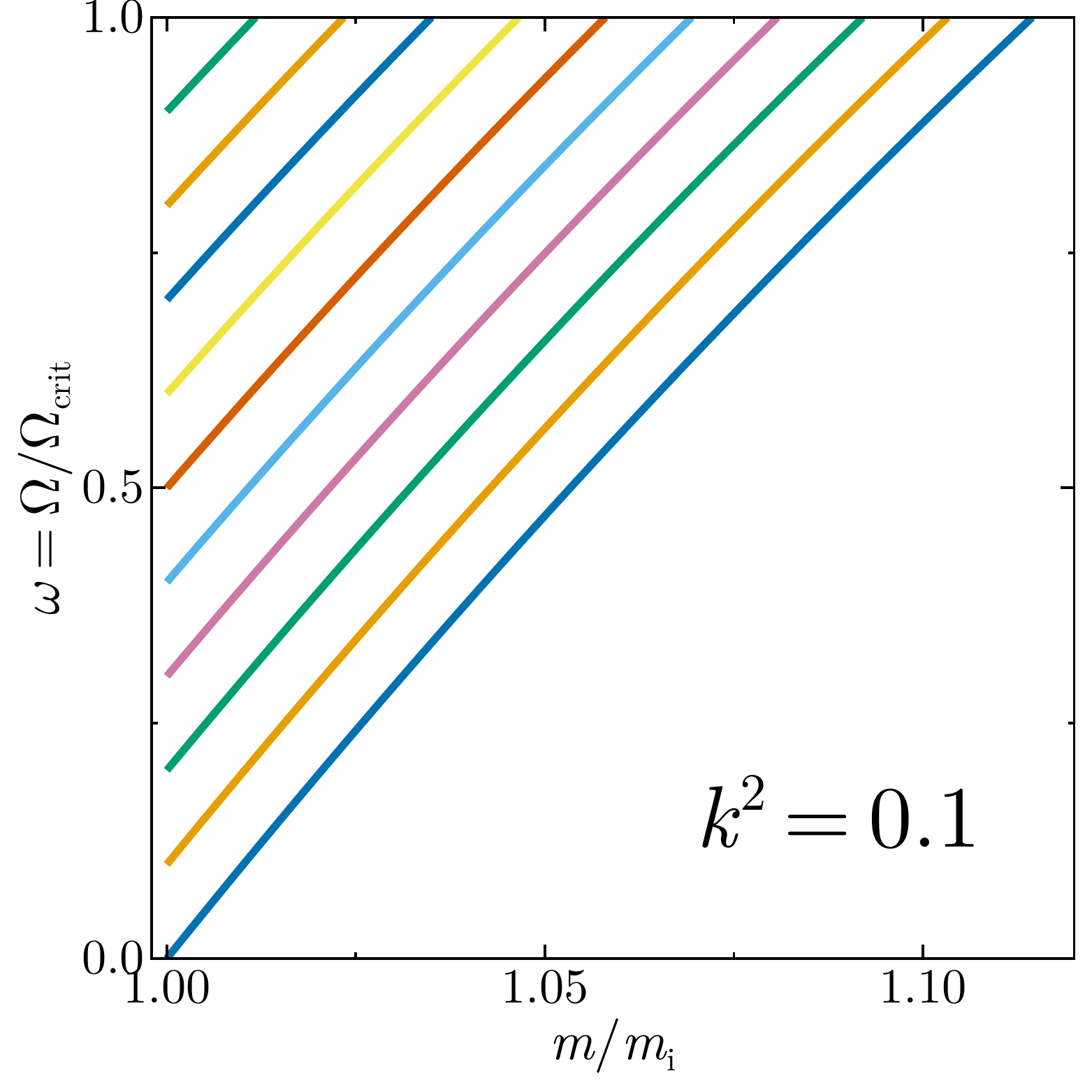} 
\end{center}
\vspace{-0.5cm}
    \caption{Spin-up of an accreting star in a binary system as a function of its mass following the model by \citet{Packet1981}. Each line corresponds to a different choice for the initial ratio of the rotational frequency to the critical rotational frequency of the star, taking a value of $k^2=0.1$ (where $kR$ is the gyration radius), which is representative of stars on the main sequence.}
    \label{fig:spinup}
\end{wrapfigure}

The identification of an observed star as a product of binary accretion is, however, not always straightforward. Blue stragglers in clusters can be identified by their location in a color-magnitude diagram, but if found in isolation their age cannot be assessed in this way. Moreover, even in a cluster rejuvenated stars can still be present below the turnoff, which were dubbed as "blue lurkers" by \cite{Leiner+2019}. The lack of an observed binary companion, is also not sufficient to exclude prior interaction. The evolutionary sequence shown in Figure \ref{fig:vdh} shows a system that remains bound after the formation of a compact object, but supernovae can inject large amounts of momentum into a newly formed compact object, unbinding the binary system. Observational biases can also prevent the identification of binaries with mass ratios significantly away from unity \citep{Sana2017}. Owing to these two effects, excluding known binaries from an observed sample can actually increase the fraction of binary products, as the most easily identifiable binaries are often in a pre-interaction stage \citep{deMink+2014}. As mass accretion brings a significant amount of angular momentum into a star, rapid rotation as well as the presence of material exhibiting signatures of nuclear processing (in particular nitrogen enrichment) are also indicative of former interaction. As a consequence of rapid rotation, stars can exhibit the Be phenomenon, where emission features produced in a decretion disk are visible in their spectra \citep{Rivinius+2013}. The joint presence of high rotation and enrichment in CNO processed material is, however, degenerate with the expectations from single rapidly rotating stars (see \citealt{Langer2012,Marchant2024} for a discussion).

% Packet
The accretion of angular momentum brings however an important problem. If accretion happens through the formation of an accretion disk, then the accreted material carries a large amount of angular momentum, quickly spinning up the star. A simple model to quantify how fast spinup happens was developed by \citet{Packet1981}, were during the accretion phase the star is assumed to have a constant radius $R$ as well as a constant gyration radius $kR$. In this case, the spin angular momentum of the star and its time derivative are given by
\begin{eqnarray}
    J = \Omega k^2 m R^2 \rightarrow \frac{\dot{J}}{J}=\frac{\dot{\Omega}}{\Omega} + \frac{\dot{m}}{m},
\end{eqnarray}
were $J$ is the spin angular momentum of the accretor, $m$ is its mass and $\Omega$ its rotational frequency, assumed constant through the star. To assess how fast the star is rotating, we can consider the ratio of its rotational frequency to the Keplerian rate at its surface,
\begin{eqnarray}
    \omega = \frac{\Omega}{\sqrt{Gm/R^3}} \rightarrow \frac{\dot{\omega}}{\omega}=\frac{\dot{\Omega}}{\Omega}-\frac{1}{2}\frac{\dot{m}}{m}.
\end{eqnarray}
The rate of change in spin angular momentum of the accretor is given by the assumption of accretion through a Keplerian disk,
\begin{eqnarray}
    \dot{J} = \dot{m}\sqrt{GmR}\rightarrow \frac{\dot{J}}{J}=\frac{1}{\omega k^2}\frac{\dot{m}}{m}.
\end{eqnarray}
These three equations can be combined into a single differential equation for $\omega(m)$ which has an analytical solution,
\begin{eqnarray}
    \frac{\mathrm{d}\omega}{\mathrm{d}m}=\frac{1}{m}\left(\frac{1}{k^2}-\frac{3\omega}{2}\right)\rightarrow \omega(m/m_\mathrm{i}) = \left(\omega_\mathrm{i}-\frac{2}{3k^2}\right)\left(\frac{m}{m_\mathrm{i}}\right)^{-3/2}+\frac{2}{3k^2}, \label{equ:spin}
\end{eqnarray}
where $\omega_\mathrm{i}$ and $m_\mathrm{i}$ are the values at the beginning of the accretion phase.

The result of Equation (\ref{equ:spin}) is shown in Figure \ref{fig:spinup} for different initial values of $\omega_i$ and a value of $k^2=0.1$, representative of stars on the main sequence. Even for the case of an initially non-rotating accretor, an increase in mass of only $\sim10\%$ is enough to drive it to critical rotation. The simplifications made in this model actually alleviate the problem compared to reality. The energy injected through accretion can lead to an expansion of the outermost layers of the star, leading to a decrease in the critical rotation frequency and an increase in the angular momentum accreted through the disk. Moreover, angular momentum is not transported instantaneously through the star, and a differentially rotating structure, with a faster surface rotation, can reach critical rotation faster.

The question of what happens once the accretor reaches critical rotation is a significant uncertainty in binary evolution models, as further accretion would require a mechanism to remove angular momentum from the star. One working assumption is to consider that once critical rotation is achieved further mass being transferred is ejected from the binary \citep{Langer1997}. The efficiency of accretion is then mediated by the strength of tides, leading to binaries with initially shorter periods undergoing more efficient mass transfer stages \citep{Sen+2022,Rocha+2024}. Observations of various post-interaction products indicate however that mass transfer can be very efficient \citep{Schootemeijer+2018, Vinciguerra+2020, Bodensteiner+2020}. An alternative mechanism that allows for accretion to proceed is that once critical rotation is reached the surplus angular momentum is transported outwards through the disk and coupled back to the orbit \citep{PophamNarayan1991,Paczynski1991}. The solution to this problem will likely require the use of multi-D hydrodynamical simulations coupled with careful comparison to observed binary systems, both actively undergoing mass transfer as well as post-detachment. Of particular interest are systems undergoing thermal timescale mass transfer, such as $\beta$ Lyrae \citep{Zhao+2008,Mourard+2018}.

\section{End-stages of binary evolution}
The impact of binary interactions affects all subsequent stages of evolution of both components, leading to various outcomes that differ from single star evolution. Properly reviewing all these processes is beyond the scope of this paper, but in the following I discuss some key aspects related to end-stages of binary evolution, where by end-stage I refer to cases where the binary ceases to exist as such, or one of the stars produces a compact object. For extensive recent reviews the reader is directed to \citet{Chen+2024} and \citet{Marchant2024}.

\subsection{Mergers}
As mentioned, various situations can lead to dynamical instability and a stellar merger, including mass loss from the outer Lagrangian points, the Darwin instability and common envelope evolution. Stellar mergers can lead to bright transient events known as luminous red novae (see \citealt{Pastorello+2019} and references within), with the particular case of V1309 Sco that was confidently associated with a stellar coalescence \citep{Tylenda+2011}.

The properties of a merged star can vary depending on the nature of the merging components. Mergers of main-sequence stars can lead to rejuvenated main-sequence objects, which can also appear as blue stragglers in stellar clusters \citep{Mateo+1990}. On the opposite extreme mergers of compact objects with stars (such as a possible merger between the neutron star and the Be star in the subsequent evolution of Figure \ref{fig:vdh}) has been suggested to lead to an exotic type of star known as a Thorne-Żytkow object, where the neutron star sinks to the core of the merger with shell burning happening on top of it \citep{ThorneZytkow1975,Farmer+2023}. It is commonly expected that the merger process deposits a significant amount of the orbital angular momentum in the remnant leading to rapid rotation, which has been suggested as an explanation for the large projected rotational velocity measured for Betelgeuse (e.g. \citealt{Shiber+2024}, although see \citealt{JingZe+2024} for a counterargument). Using 3D simulations coupled with 1D calculations to follow the evolution of the merger product, \citet{Schneider+2019} has argued that the thermal readjustement and mass loss of the star post-merger leads to the formation of a slowly spinning star instead.

As stellar mergers can lead to significant differential rotation, it has been suggested they can be the source of stellar magnetism in early type stars \citep{Ferrario+2009}. Contrary to low mass stars like the sun which keep a magnetic field in their surfaces through a sustained dynamo process in their convective envelope, early type stars have stable radiative envelopes, and those that are observed to be magnetic are considered to have "fossil" fields inherited from an earlier evolutionary stage \citep{DonatiLandstreet2009}. The merger scenario for magnetic field generation in main sequence stars has been shown in action through 3D magnetohydrodynamic simulations, showing indeed that large scale fields can be produced \citep{Schneider+2019}. Very strong observational evidence of this process was recently provided by \citet{Frost+2024}, who showed that the magnetic star HD 14893 is a rejuvenated object. Mergers have also been argued to be the origin of some magnetic white dwarfs \citep{Tout+2008, GarciaBerro+2012} and subdwarf stars \citep{Pelisoli+2022}. The highest magnetic field measured in a non-degenerate star is that of the $\sim 2M_\odot$ stripped star HD 45166, which has been argued to be the product of the merger of two post-main sequence stars \citep{Shenar+2023}. Interestingly, even though both HD 14893 and HD 45166 are though to be the product of a stellar merger, they are both observed to be in long period binaries. The current binary companion is understood to have actually been the outer component of a hierarchical triple before the merger event.
\subsection{Supernovae}
The occurrence of binary interactions can also significantly affect the types of supernovae that we observe. Akin to the situation with blue stragglers, rejuvenated stars can lead to supernovae from old stellar populations where all single massive stars have reached core-collapse \citep{Zapartas+2017}. But structural changes can also significantly alter the properties of supernova progenitors. Of particular historical interest to binary evolution is the supernova 1987a, which occurred in the nearby Large Magellanic Cloud. Owing to its proximity, this was the first core-collapse event for which there was a clear detection of a progenitor, identified as a blue-supergiant star in pre-explosion archival images \citep{Panagia+1987}. This provided a direct comparison point against final evolutionary stages predicted by stellar evolution models. This was not a regular prediction in stellar evolution calculations, so it was suggested that binary evolution could be the cause behind it (\citealt{PodsiadlowskiJoss1989}, see \citealt{MenonHeger2017} for more recent simulations). Back at that time binary evolution was not considered a very important aspect of stellar evolution, and this supernova placed some significant momentum into the field. In this particular binary formation scenario, either accretion or a merger event can lead to a structure with overmassive hydrogen envelopes as compared to single star evolution, which can result in core-helium burning and even later stages being carried out as a blue supergiant without undergoing a red supergiant phase \citep{Justham+2014}.

The formation of hydrogen-poor supernovae can also be enhanced by binary interactions. Core collapse supernovae are categorized into types II and I depending on on the presence or absence of hydrogen features in their spectra. Type I supernovae are further categorized into types Ia which exhibit silicon lines indicative of a thermonuclear explosion rather than core-collapse, or Ib/Ic (depending on the presence or abscence of helium features). The different binary processes that lead to these different supernovae types are extensively discussed by \cite{Podsiadlowski+1992}. In binary systems, donor stars that would not be massive enough to lose their outer envelopes through stellar winds can still be stripped of their hydrogen envelopes and produce type Ib supernovae \citep{Yoon+2010}. Binary stripped stars can even retain a significant amount of hydrogen up to core-collapse, leading to type IIb supernovae, a particular type that transitions between type II and Ib \citep{Yoon+2017,Sravan+2019}. 

Binary interactions are also key to the production of type Ia supernovae, which correspond to thermonuclear explosions of white dwarfs which leave no remnants \citep{HoyleFowler1960}. Two main scenarios are thought to contribute: the single-degenerate case where a white dwarf accretes from a non-degenerate companion until it exceeds its Chandrasekhar limit \cite{WhelanIben1973} and the double degenerate scenario, where two merging binary white dwarfs coalesce due to emission of gravitational waves and initiate a thermonuclear explosion \citep{Webbink1984,IbenTutukov1984}. As type Ia are an important standard candle used in cosmology \citep{Riess+1998}, understanding their origin is of particular importance. Recent reviews discussing the evidence for and against each channel are those of \citet{Maoz+2014} and \citet{Liu+2023}.

\subsection{Single-degenerate binaries}
A diverse set of binary products is composed of a component that has reached an evolutionary end point (be it a white-dwarf, neutron star or black hole) with a non-degenerate companion. An excellent overview of their formation and evolution is provided in the textboox by \cite{Tauris2023}. Such systems are referred to as single-degenerate binaries and when part of an interacting binary can also be observed as X-ray binaries. Systems composed of a white dwarf accretor and a mass transferring companion are known as cataclysmic variables, and depending on the mass of the white dwarf, the accretion rate and composition of accreted material, these systems can undergo bright nova eruptions (see \citealt{Chomiuk+2021} and references within).

Neutron stars in binary systems can evolve very differently depending on their companions. In high mass X-ray binaries (where the companion is a massive star), a common type of system is a Be X-ray binary as illustrated in the last step of Figure \ref{fig:vdh}. Neutron stars in Be X-ray binaries can accrete material from the decretion disk of the Be star without the need for Roche-lobe overflow \citep{Reig2011}. Accretion rates through this phase are low, so the neutron star is not expected to grow, and even in later stages where the donor fills its Roche lobe mass transfer is expected to be either unstable (possibly leading to a Thorne-Żytkow object), or with mass transfer rates well above the Eddington limit for accretion of the neutron star (at which point the luminosity becomes so high that radiation pressure prevents further accretion). In cases were the companion to the neutron star is a low mass star, significant mass can be transferred leading to growth of the neutron star mass and the possible formation of a millisecond pulsar \citep{TaurisSavonije1999}. With orbital evolution driven by magnetic braking and gravitational wave radiation, the masses of the donor stars in low mass X-ray binaries can be significantly reduced, and as such they are referred to as black widow or redback systems (see \citealt{Chen+2013} and references within).

Black hole X-ray binaries represent the main way in which stellar mass black holes had been detected and weighted prior to the detection of gravitational wave sources. Within our Galaxy, the largest black hole mass measured in an X-ray binary is that of Cygnus X-1 at $21.2\pm 2.2M_\odot$ \citep{Miller-Jones2021}. Single-degenerate binaries can also be detected in situations where there is no significant mass transfer and the compact object is effectively just a dark point mass. With a mass of $32.7\pm0.82M_\odot$ the record holder for most massive stellar mass black hole detected with electromagnetic observations is one such non-interacting system, identified through astrometric observations of the motion of its companion star by the Gaia mission \citep{Panuzzo+2024}.
\subsection{Gravitational wave sources}

%detections, future prospects
Starting from 2015, gravitational wave astrophysics became a new window to study stellar evolution and binary interactions in particular. This discovery was the culmination of significant technological developments allowing for extremely precise measurements of the displacement of test masses in kilometer-wide interferomenters (see \citealt{Bond+2016} for an overview), with the two detectors of the Laser Interferometer Gravitational wave Observatory in the USA (LIGO, \citealt{aLIGO}) and Virgo in Italy \citep{aVirgo} having reported detections so far. The first source discovered, GW1510914, consisted of the merger of two binary black holes with masses of $\sim 35M_\odot+29M_\odot$ \citep{GW150914}, well beyond the dynamical masses measured with electromagnetism in black hole binaries. In less than a decade, the rate of discovery when the detectors are in operation has increased from one event per month, to almost one event per day, with the current catalogue reaching almost a hundred sources and including both merging binary neutron stars and mergers between black holes and neutron stars \citep{GWTC3}. Among notable events, GW170817 consisted of a merger between two neutron stars for which an optical transient as well as an associated gamma-ray burst were detected \citep{GW170817}. This provided direct proof of the association of gamma-ray bursts with compact object mergers, and the direct measurement of the redshift allowed for its use as a standard candle for cosmology \citep{GW170817_cosmo}.
\begin{figure*}
\begin{center}
\includegraphics[width=\textwidth]{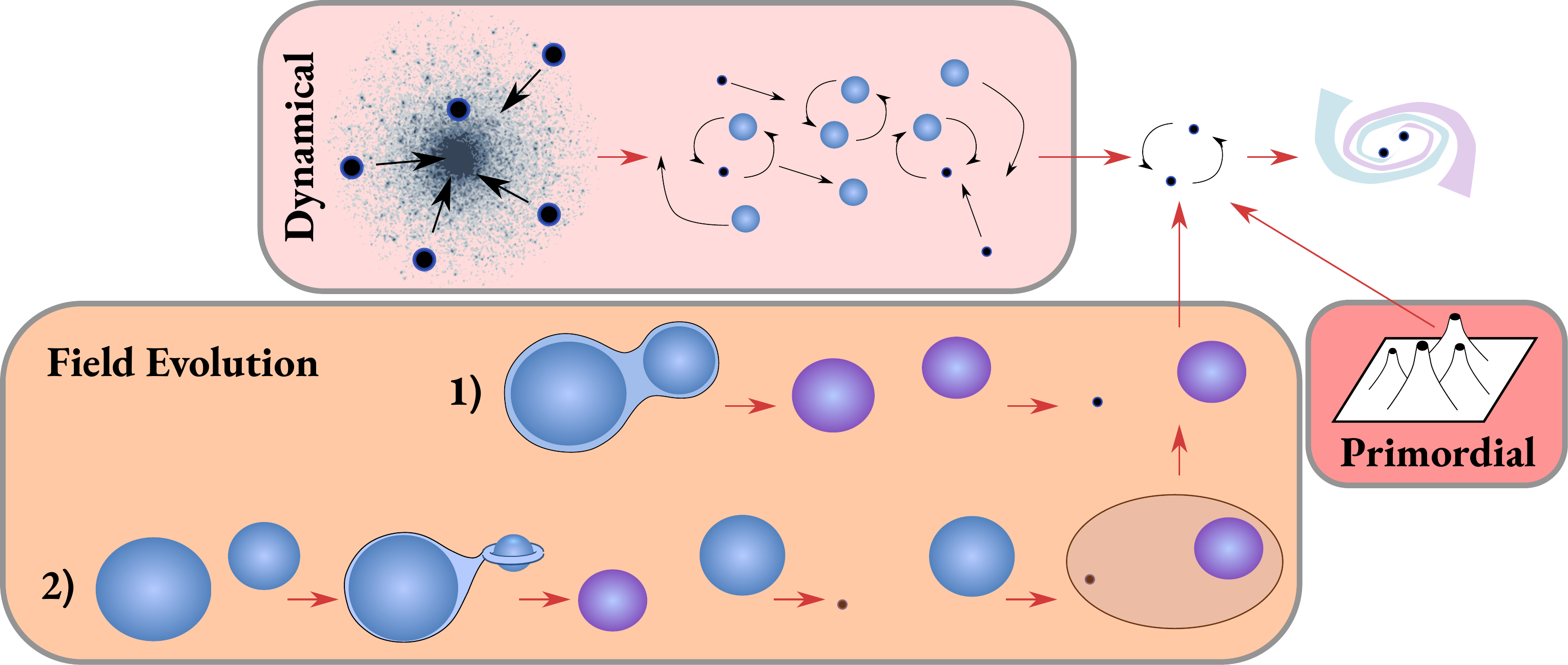} 
\end{center}
    \caption{Examples of proposed evolutionary channels to form merging binary black holes.}
    \label{fig:gw}
\end{figure*}

Improvements to the current generation of detectors, as well as projects for 3rd generation ground-based interferometers will allow for precise constraints on the population properties of merging compact objects across cosmic time \citep{Baibhav+2019,HallEvans2019}. One critical limitation of ground-based detectors, however, is that seismic noise drastically limits their sensitivity below 1 Hz, requiring different techniques to probe the low-frequency gravitational wave sky. One such project is the Laser Interferometer Space Antenna (LISA, planned launch mid-2030), a constellation of three spacecraft in a triangular configuration separated by about 2.5 million kilometers \citep{LISA_wp}. This will open up the gravitational frequency spectrum to frequencies down to $10^{-5}$ Hz, allowing for the detection of lower mass gravitational wave sources such as short period binary white dwarfs. Through electromagnetic observations there are already known systems (referred to as verification binaries) that should be detected by LISA if it operates at its expected sensitivity (e.g. \citealt{Kupfer+2018}).

%requirements for a merger in terms of separation
The mechanism through which the observed gravitational wave sources were produced is, however, uncertain. Key to understand their formation is that any evolutionary process to explain them needs to result in compact object pairs at a sufficiently small separation for gravitational wave radiation to be significant. The time it takes for a circular binary with a separation $a$ and masses $m_1$ and $m_2$ to merge due to gravitational wave radiation is \citep{Peters1964}
\begin{eqnarray}
    t_\mathrm{merge} = \frac{5}{256}\frac{c^5}{G^3}\frac{a^4}{m_1 m_2 (m_1+m_2)}.
\end{eqnarray}
Using this one can consider the required orbital separation for a binary with given masses to merge within the lifetime of the universe of 13.8 Gyrs \citep{Planck_results}. Considering for instance two white dwarfs with masses of $0.5M_\odot$, orbital separations $<2.2R_\odot$ are required. Taking two black holes of $30M_\odot$, the separation must be $<48R_\odot$. If such sources are formed in binary systems, it necessarily involves binary interaction, as stars can expand to much larger radii. In order to form such compact object binaries, generally a process that shrinks the orbit is required. The classical formation scenario involves a common-envelope phase where the outer envelope is ejected at the cost of orbital energy, leading to a shorter period binary. This process was initially proposed to explain the existence of short period cataclysmic variables \citep{Paczynski1976}, but it applies similarly to the formation of merging binary neutron stars and black holes (see \citealt{Bavera+2021,Gallegos-Garcia+2023} for recent results). An alternative evolutionary pathway is that of chemically-homogeneous evolution, where two massive stars are born in a very close orbit (possibly in contact) and retain their short separation until the formation of a binary black hole \citep{MandelDeMink2016,Marchant+2016}. In particular for the case of merging binary black holes, important issues have been shown for predictions of the common envelope channel \citep{Klencki+2021}, and the dominant evolutionary channel could be hardening through stable mass transfer \citep{vandenheuvel+2017,Marchant+2021,Gallegos-Garcia2021, Picco+2024}.

%variety of formation channels
Ideally, with an increasing number of gravitational wave observations it would be possible to distinguish between multiple formation channels. However, predicting the current population of gravitational wave sources requires a full model of the Universe, since owing to possible large delay times sources observed to merge at redshift zero could have been formed in the early Universe. This is compounded by present uncertainties in binary evolution modelling, and also by the presence of different channels (illustrated in Figure \ref{fig:gw}) that do not require binary interaction (see \citealt{MandelFarmer2022} for a recent review). Even if one has a set of well understood predictions from certain evolutionary processes, population inference done in the observed sample can lead to incorrect results owing to contributions from unknown or unnacounted channels \citep{Cheng+2023}. It is then key to take advantage of electromagnetic observations of intermediate evolutionary stages to anchor theoretical models and assess what processes actually contribute to the observed sample.

\section{Conclusion} 

Binary evolution results in a rich set of interaction processes and outcomes that are not possible for stars evolving in isolation. This comes, however, with increased complexity in their evolution and significant modeling uncertainties. Interactions in multiple systems are also not limited to two components, with the study of interactions in triple star systems also gaining significant attention in recent years (see \citealt{Kummer+2023} for an overview). Theoretical predictions of the evolution of binary systems use multiple techniques, including 1D approximations, full 3D hydrodynamics as well as synthetic synthesis of entire populations in order to compare with observations, often required in combination to study a particular evolutionary stage. Owing to current and future surveys, the data available to constrain our theories will increase dramatically in coming years. This includes astrometric and photometric observations of binary systems from the Gaia mission \citep{DR3binaries}, transient surveys such as the Zwicky Transient Facility \citep{ZTF} and the Vera Rubin observatory \citep{Rubin2019}, and spectroscopic surveys such as BLOeM \citep{Shenar+2024}. The ever increasing sensitivity of gravitational wave detectors will also provide a unique view of binary evolution at late stages, specially in the following decade as next generation detectors begin operation and LISA is launched. These rich datasets with well understood biases will provide key constraints to our evolutionary models, allowing for clear comparison to theoretical predictions and to find solutions to long-standing problems in stellar evolution theory.

\section{Acknowledgements}
PM acknowledges support from the FWO senior postdoctoral fellowship number 12ZY523N.

\bibliographystyle{Harvard}
%\bibliography{binaries}
\begin{thebibliography*}{133}
\providecommand{\bibtype}[1]{}
\providecommand{\natexlab}[1]{#1}
{\catcode`\|=0\catcode`\#=12\catcode`\@=11\catcode`\\=12
|immediate|write|@auxout{\expandafter\ifx\csname natexlab\endcsname\relax\gdef\natexlab#1{#1}\fi}}
\renewcommand{\url}[1]{{\tt #1}}
\providecommand{\urlprefix}{URL }
\expandafter\ifx\csname urlstyle\endcsname\relax
  \providecommand{\doi}[1]{doi:\discretionary{}{}{}#1}\else
  \providecommand{\doi}{doi:\discretionary{}{}{}\begingroup \urlstyle{rm}\Url}\fi
\providecommand{\bibinfo}[2]{#2}
\providecommand{\eprint}[2][]{\url{#2}}

\bibtype{Article}%
\bibitem[{Aasi} et al.(2015)]{aLIGO}
\bibinfo{author}{{Aasi} J}, \bibinfo{author}{{Abbott} BP}, \bibinfo{author}{{Abbott}} and  \bibinfo{author}{et. al.} (\bibinfo{year}{2015}), \bibinfo{month}{Apr.}
\bibinfo{title}{{Advanced LIGO}}.
\bibinfo{journal}{{\em Classical and Quantum Gravity}} \bibinfo{volume}{32} (\bibinfo{number}{7}), \bibinfo{eid}{074001}. \bibinfo{doi}{\doi{10.1088/0264-9381/32/7/074001}}.
\eprint{1411.4547}.

\bibtype{Article}%
\bibitem[{Abbott} et al.(2016)]{GW150914}
\bibinfo{author}{{Abbott} BP}, \bibinfo{author}{{Abbott} R}, \bibinfo{author}{{Abbott}} and  \bibinfo{author}{et. al.} (\bibinfo{year}{2016}), \bibinfo{month}{Feb.}
\bibinfo{title}{{Observation of Gravitational Waves from a Binary Black Hole Merger}}.
\bibinfo{journal}{{\em \prl}} \bibinfo{volume}{116} (\bibinfo{number}{6}), \bibinfo{eid}{061102}. \bibinfo{doi}{\doi{10.1103/PhysRevLett.116.061102}}.
\eprint{1602.03837}.

\bibtype{Article}%
\bibitem[{Abbott} et al.(2017{\natexlab{a}})]{GW170817_cosmo}
\bibinfo{author}{{Abbott} BP}, \bibinfo{author}{{Abbott} R}, \bibinfo{author}{{Abbott}} and  \bibinfo{author}{et. al.} (\bibinfo{year}{2017}{\natexlab{a}}), \bibinfo{month}{Nov.}
\bibinfo{title}{{A gravitational-wave standard siren measurement of the Hubble constant}}.
\bibinfo{journal}{{\em \nat}} \bibinfo{volume}{551} (\bibinfo{number}{7678}): \bibinfo{pages}{85--88}. \bibinfo{doi}{\doi{10.1038/nature24471}}.
\eprint{1710.05835}.

\bibtype{Article}%
\bibitem[{Abbott} et al.(2017{\natexlab{b}})]{GW170817}
\bibinfo{author}{{Abbott} BP}, \bibinfo{author}{{Abbott} R}, \bibinfo{author}{{Abbott}} and  \bibinfo{author}{et. al.} (\bibinfo{year}{2017}{\natexlab{b}}), \bibinfo{month}{Oct.}
\bibinfo{title}{{GW170817: Observation of Gravitational Waves from a Binary Neutron Star Inspiral}}.
\bibinfo{journal}{{\em \prl}} \bibinfo{volume}{119} (\bibinfo{number}{16}), \bibinfo{eid}{161101}. \bibinfo{doi}{\doi{10.1103/PhysRevLett.119.161101}}.
\eprint{1710.05832}.

\bibtype{Article}%
\bibitem[{Abbott} et al.(2023)]{GWTC3}
\bibinfo{author}{{Abbott} R}, \bibinfo{author}{{Abbott} TD}, \bibinfo{author}{{Acernese} F} and  \bibinfo{author}{et~al.} (\bibinfo{year}{2023}), \bibinfo{month}{Oct.}
\bibinfo{title}{{GWTC-3: Compact Binary Coalescences Observed by LIGO and Virgo during the Second Part of the Third Observing Run}}.
\bibinfo{journal}{{\em Physical Review X}} \bibinfo{volume}{13} (\bibinfo{number}{4}), \bibinfo{eid}{041039}. \bibinfo{doi}{\doi{10.1103/PhysRevX.13.041039}}.
\eprint{2111.03606}.

\bibtype{Article}%
\bibitem[{Acernese} et al.(2015)]{aVirgo}
\bibinfo{author}{{Acernese} F}, \bibinfo{author}{{Agathos} M}, \bibinfo{author}{{Agatsuma}} and  \bibinfo{author}{et. al.} (\bibinfo{year}{2015}), \bibinfo{month}{Jan.}
\bibinfo{title}{{Advanced Virgo: a second-generation interferometric gravitational wave detector}}.
\bibinfo{journal}{{\em Classical and Quantum Gravity}} \bibinfo{volume}{32} (\bibinfo{number}{2}), \bibinfo{eid}{024001}. \bibinfo{doi}{\doi{10.1088/0264-9381/32/2/024001}}.
\eprint{1408.3978}.

\bibtype{Article}%
\bibitem[{Agertz} et al.(2013)]{Agertz+2013}
\bibinfo{author}{{Agertz} O}, \bibinfo{author}{{Kravtsov} AV}, \bibinfo{author}{{Leitner} SN} and  \bibinfo{author}{{Gnedin} NY} (\bibinfo{year}{2013}), \bibinfo{month}{Jun.}
\bibinfo{title}{{Toward a Complete Accounting of Energy and Momentum from Stellar Feedback in Galaxy Formation Simulations}}.
\bibinfo{journal}{{\em \apj}} \bibinfo{volume}{770} (\bibinfo{number}{1}), \bibinfo{eid}{25}. \bibinfo{doi}{\doi{10.1088/0004-637X/770/1/25}}.
\eprint{1210.4957}.

\bibtype{Article}%
\bibitem[{Aghanim} et al.(2020)]{Planck_results}
\bibinfo{author}{{Aghanim} N}, \bibinfo{author}{{Akrami} Y}, \bibinfo{author}{{Ashdown}} and  \bibinfo{author}{et. al.} (\bibinfo{year}{2020}), \bibinfo{month}{Sep.}
\bibinfo{title}{{Planck 2018 results. VI. Cosmological parameters}}.
\bibinfo{journal}{{\em \aap}} \bibinfo{volume}{641}, \bibinfo{eid}{A6}. \bibinfo{doi}{\doi{10.1051/0004-6361/201833910}}.
\eprint{1807.06209}.

\bibtype{Article}%
\bibitem[{Althaus} et al.(2001)]{Althaus+2001}
\bibinfo{author}{{Althaus} LG}, \bibinfo{author}{{Serenelli} AM} and  \bibinfo{author}{{Benvenuto} OG} (\bibinfo{year}{2001}), \bibinfo{month}{Jun.}
\bibinfo{title}{{Formation and Evolution of a 0.242 M$_{solar}$ Helium White Dwarf in the Presence of Element Diffusion}}.
\bibinfo{journal}{{\em \apj}} \bibinfo{volume}{554} (\bibinfo{number}{2}): \bibinfo{pages}{1110--1117}. \bibinfo{doi}{\doi{10.1086/321414}}.
\eprint{astro-ph/0010169}.

\bibtype{Article}%
\bibitem[{Amaro-Seoane} et al.(2023)]{LISA_wp}
\bibinfo{author}{{Amaro-Seoane} P}, \bibinfo{author}{{Andrews} J}, \bibinfo{author}{{Arca Sedda} M} and  \bibinfo{author}{et. al.} (\bibinfo{year}{2023}), \bibinfo{month}{Dec.}
\bibinfo{title}{{Astrophysics with the Laser Interferometer Space Antenna}}.
\bibinfo{journal}{{\em Living Reviews in Relativity}} \bibinfo{volume}{26} (\bibinfo{number}{1}), \bibinfo{eid}{2}. \bibinfo{doi}{\doi{10.1007/s41114-022-00041-y}}.
\eprint{2203.06016}.

\bibtype{Article}%
\bibitem[{Arenou} et al.(2023)]{DR3binaries}
\bibinfo{author}{{Arenou} F}, \bibinfo{author}{{Babusiaux} C}, \bibinfo{author}{{Barstow} MA} and  \bibinfo{author}{et~al.} (\bibinfo{year}{2023}), \bibinfo{month}{Jun.}
\bibinfo{title}{{Gaia Data Release 3. Stellar multiplicity, a teaser for the hidden treasure}}.
\bibinfo{journal}{{\em \aap}} \bibinfo{volume}{674}, \bibinfo{eid}{A34}. \bibinfo{doi}{\doi{10.1051/0004-6361/202243782}}.
\eprint{2206.05595}.

\bibtype{Article}%
\bibitem[{Baibhav} et al.(2019)]{Baibhav+2019}
\bibinfo{author}{{Baibhav} V}, \bibinfo{author}{{Berti} E}, \bibinfo{author}{{Gerosa} D}, \bibinfo{author}{{Mapelli} M}, \bibinfo{author}{{Giacobbo} N}, \bibinfo{author}{{Bouffanais} Y} and  \bibinfo{author}{{Di Carlo} UN} (\bibinfo{year}{2019}), \bibinfo{month}{Sep.}
\bibinfo{title}{{Gravitational-wave detection rates for compact binaries formed in isolation: LIGO/Virgo O3 and beyond}}.
\bibinfo{journal}{{\em \prd}} \bibinfo{volume}{100} (\bibinfo{number}{6}), \bibinfo{eid}{064060}. \bibinfo{doi}{\doi{10.1103/PhysRevD.100.064060}}.
\eprint{1906.04197}.

\bibtype{Article}%
\bibitem[{Baron} et al.(2012)]{Baron+2012}
\bibinfo{author}{{Baron} F}, \bibinfo{author}{{Monnier} JD}, \bibinfo{author}{{Pedretti} E}, \bibinfo{author}{{Zhao} M}, \bibinfo{author}{{Schaefer} G}, \bibinfo{author}{{Parks} R}, \bibinfo{author}{{Che} X}, \bibinfo{author}{{Thureau} N}, \bibinfo{author}{{ten Brummelaar} TA}, \bibinfo{author}{{McAlister} HA}, \bibinfo{author}{{Ridgway} ST}, \bibinfo{author}{{Farrington} C}, \bibinfo{author}{{Sturmann} J}, \bibinfo{author}{{Sturmann} L} and  \bibinfo{author}{{Turner} N} (\bibinfo{year}{2012}), \bibinfo{month}{Jun.}
\bibinfo{title}{{Imaging the Algol Triple System in the H Band with the CHARA Interferometer}}.
\bibinfo{journal}{{\em \apj}} \bibinfo{volume}{752} (\bibinfo{number}{1}), \bibinfo{eid}{20}. \bibinfo{doi}{\doi{10.1088/0004-637X/752/1/20}}.
\eprint{1205.0754}.

\bibtype{Article}%
\bibitem[{Bavera} et al.(2021)]{Bavera+2021}
\bibinfo{author}{{Bavera} SS}, \bibinfo{author}{{Fragos} T}, \bibinfo{author}{{Zevin} M}, \bibinfo{author}{{Berry} CPL}, \bibinfo{author}{{Marchant} P}, \bibinfo{author}{{Andrews} JJ}, \bibinfo{author}{{Coughlin} S}, \bibinfo{author}{{Dotter} A}, \bibinfo{author}{{Kovlakas} K}, \bibinfo{author}{{Misra} D}, \bibinfo{author}{{Serra-Perez} JG}, \bibinfo{author}{{Qin} Y}, \bibinfo{author}{{Rocha} KA}, \bibinfo{author}{{Rom{\'a}n-Garza} J}, \bibinfo{author}{{Tran} NH} and  \bibinfo{author}{{Zapartas} E} (\bibinfo{year}{2021}), \bibinfo{month}{Mar.}
\bibinfo{title}{{The impact of mass-transfer physics on the observable properties of field binary black hole populations}}.
\bibinfo{journal}{{\em \aap}} \bibinfo{volume}{647}, \bibinfo{eid}{A153}. \bibinfo{doi}{\doi{10.1051/0004-6361/202039804}}.
\eprint{2010.16333}.

\bibtype{Article}%
\bibitem[{Bellm} et al.(2019)]{ZTF}
\bibinfo{author}{{Bellm} EC}, \bibinfo{author}{{Kulkarni} SR}, \bibinfo{author}{{Graham} MJ},  and  \bibinfo{author}{et. al.} (\bibinfo{year}{2019}), \bibinfo{month}{Jan.}
\bibinfo{title}{{The Zwicky Transient Facility: System Overview, Performance, and First Results}}.
\bibinfo{journal}{{\em \pasp}} \bibinfo{volume}{131} (\bibinfo{number}{995}): \bibinfo{pages}{018002}. \bibinfo{doi}{\doi{10.1088/1538-3873/aaecbe}}.
\eprint{1902.01932}.

\bibtype{Article}%
\bibitem[{Bodensteiner} et al.(2020)]{Bodensteiner+2020}
\bibinfo{author}{{Bodensteiner} J}, \bibinfo{author}{{Shenar} T}, \bibinfo{author}{{Mahy} L}, \bibinfo{author}{{Fabry} M}, \bibinfo{author}{{Marchant} P}, \bibinfo{author}{{Abdul-Masih} M}, \bibinfo{author}{{Banyard} G}, \bibinfo{author}{{Bowman} DM}, \bibinfo{author}{{Dsilva} K}, \bibinfo{author}{{Frost} AJ}, \bibinfo{author}{{Hawcroft} C}, \bibinfo{author}{{Reggiani} M} and  \bibinfo{author}{{Sana} H} (\bibinfo{year}{2020}), \bibinfo{month}{Sep.}
\bibinfo{title}{{Is HR 6819 a triple system containing a black hole?. An alternative explanation}}.
\bibinfo{journal}{{\em \aap}} \bibinfo{volume}{641}, \bibinfo{eid}{A43}. \bibinfo{doi}{\doi{10.1051/0004-6361/202038682}}.
\eprint{2006.10770}.

\bibtype{Article}%
\bibitem[{Bond} et al.(2016)]{Bond+2016}
\bibinfo{author}{{Bond} C}, \bibinfo{author}{{Brown} D}, \bibinfo{author}{{Freise} A} and  \bibinfo{author}{{Strain} KA} (\bibinfo{year}{2016}), \bibinfo{month}{Dec.}
\bibinfo{title}{{Interferometer techniques for gravitational-wave detection}}.
\bibinfo{journal}{{\em Living Reviews in Relativity}} \bibinfo{volume}{19} (\bibinfo{number}{1}), \bibinfo{eid}{3}. \bibinfo{doi}{\doi{10.1007/s41114-016-0002-8}}.

\bibtype{Article}%
\bibitem[{Burbidge} and {Sandage}(1958)]{BurbidgeSandage1958}
\bibinfo{author}{{Burbidge} EM} and  \bibinfo{author}{{Sandage} A} (\bibinfo{year}{1958}), \bibinfo{month}{Sep.}
\bibinfo{title}{{The Color-Magnitude Diagram for the Galactic NGC 7789.}}
\bibinfo{journal}{{\em \apj}} \bibinfo{volume}{128}: \bibinfo{pages}{174}. \bibinfo{doi}{\doi{10.1086/146535}}.

\bibtype{Article}%
\bibitem[{Chen} et al.(2013)]{Chen+2013}
\bibinfo{author}{{Chen} HL}, \bibinfo{author}{{Chen} X}, \bibinfo{author}{{Tauris} TM} and  \bibinfo{author}{{Han} Z} (\bibinfo{year}{2013}), \bibinfo{month}{Sep.}
\bibinfo{title}{{Formation of Black Widows and Redbacks{\textemdash}Two Distinct Populations of Eclipsing Binary Millisecond Pulsars}}.
\bibinfo{journal}{{\em \apj}} \bibinfo{volume}{775} (\bibinfo{number}{1}), \bibinfo{eid}{27}. \bibinfo{doi}{\doi{10.1088/0004-637X/775/1/27}}.
\eprint{1308.4107}.

\bibtype{Article}%
\bibitem[{Chen} et al.(2024)]{Chen+2024}
\bibinfo{author}{{Chen} X}, \bibinfo{author}{{Liu} Z} and  \bibinfo{author}{{Han} Z} (\bibinfo{year}{2024}), \bibinfo{month}{Jan.}
\bibinfo{title}{{Binary stars in the new millennium}}.
\bibinfo{journal}{{\em Progress in Particle and Nuclear Physics}} \bibinfo{volume}{134}, \bibinfo{eid}{104083}. \bibinfo{doi}{\doi{10.1016/j.ppnp.2023.104083}}.
\eprint{2311.11454}.

\bibtype{Article}%
\bibitem[{Cheng} et al.(2023)]{Cheng+2023}
\bibinfo{author}{{Cheng} AQ}, \bibinfo{author}{{Zevin} M} and  \bibinfo{author}{{Vitale} S} (\bibinfo{year}{2023}), \bibinfo{month}{Oct.}
\bibinfo{title}{{What You Don't Know Can Hurt You: Use and Abuse of Astrophysical Models in Gravitational-wave Population Analyses}}.
\bibinfo{journal}{{\em \apj}} \bibinfo{volume}{955} (\bibinfo{number}{2}), \bibinfo{eid}{127}. \bibinfo{doi}{\doi{10.3847/1538-4357/aced98}}.
\eprint{2307.03129}.

\bibtype{Article}%
\bibitem[{Chomiuk} et al.(2021)]{Chomiuk+2021}
\bibinfo{author}{{Chomiuk} L}, \bibinfo{author}{{Metzger} BD} and  \bibinfo{author}{{Shen} KJ} (\bibinfo{year}{2021}), \bibinfo{month}{Sep.}
\bibinfo{title}{{New Insights into Classical Novae}}.
\bibinfo{journal}{{\em \araa}} \bibinfo{volume}{59}: \bibinfo{pages}{391--444}. \bibinfo{doi}{\doi{10.1146/annurev-astro-112420-114502}}.
\eprint{2011.08751}.

\bibtype{Article}%
\bibitem[{Colpi} et al.(2024)]{Colpi+2024}
\bibinfo{author}{{Colpi} M}, \bibinfo{author}{{Danzmann} K}, \bibinfo{author}{{Hewitson} M} and  \bibinfo{author}{et~al.} (\bibinfo{year}{2024}), \bibinfo{month}{Feb.}
\bibinfo{journal}{{\em arXiv e-prints}} , \bibinfo{eid}{arXiv:2402.07571}\bibinfo{doi}{\doi{10.48550/arXiv.2402.07571}}.
\eprint{2402.07571}.

\bibtype{Article}%
\bibitem[{Crawford}(1955)]{Crawford1955}
\bibinfo{author}{{Crawford} JA} (\bibinfo{year}{1955}), \bibinfo{month}{Jan.}
\bibinfo{title}{{On the Subgiant Components of Eclipsing Binary Systems.}}
\bibinfo{journal}{{\em \apj}} \bibinfo{volume}{121}: \bibinfo{pages}{71}. \bibinfo{doi}{\doi{10.1086/145965}}.

\bibtype{Article}%
\bibitem[{Darwin}(1879{\natexlab{a}})]{Darwin1879tides}
\bibinfo{author}{{Darwin} GH} (\bibinfo{year}{1879}{\natexlab{a}}), \bibinfo{month}{Jul.}
\bibinfo{title}{{A tidal theory of the evolution of satellites}}.
\bibinfo{journal}{{\em The Observatory}} \bibinfo{volume}{3}: \bibinfo{pages}{79--84}.

\bibtype{Article}%
\bibitem[{Darwin}(1879{\natexlab{b}})]{Darwin1879}
\bibinfo{author}{{Darwin} GH} (\bibinfo{year}{1879}{\natexlab{b}}), \bibinfo{month}{Jan.}
\bibinfo{title}{{The Determination of the Secular Effects of Tidal Friction by a Graphical Method}}.
\bibinfo{journal}{{\em Proceedings of the Royal Society of London Series I}} \bibinfo{volume}{29}: \bibinfo{pages}{168--181}.

\bibtype{Article}%
\bibitem[{de Mink} et al.(2009)]{deMink+2009}
\bibinfo{author}{{de Mink} SE}, \bibinfo{author}{{Cantiello} M}, \bibinfo{author}{{Langer} N}, \bibinfo{author}{{Pols} OR}, \bibinfo{author}{{Brott} I} and  \bibinfo{author}{{Yoon} SC} (\bibinfo{year}{2009}), \bibinfo{month}{Apr.}
\bibinfo{title}{{Rotational mixing in massive binaries. Detached short-period systems}}.
\bibinfo{journal}{{\em \aap}} \bibinfo{volume}{497} (\bibinfo{number}{1}): \bibinfo{pages}{243--253}. \bibinfo{doi}{\doi{10.1051/0004-6361/200811439}}.
\eprint{0902.1751}.

\bibtype{Article}%
\bibitem[{de Mink} et al.(2014)]{deMink+2014}
\bibinfo{author}{{de Mink} SE}, \bibinfo{author}{{Sana} H}, \bibinfo{author}{{Langer} N}, \bibinfo{author}{{Izzard} RG} and  \bibinfo{author}{{Schneider} FRN} (\bibinfo{year}{2014}), \bibinfo{month}{Feb.}
\bibinfo{title}{{The Incidence of Stellar Mergers and Mass Gainers among Massive Stars}}.
\bibinfo{journal}{{\em \apj}} \bibinfo{volume}{782} (\bibinfo{number}{1}), \bibinfo{eid}{7}. \bibinfo{doi}{\doi{10.1088/0004-637X/782/1/7}}.
\eprint{1312.3650}.

\bibtype{Article}%
\bibitem[{Donati} and {Landstreet}(2009)]{DonatiLandstreet2009}
\bibinfo{author}{{Donati} JF} and  \bibinfo{author}{{Landstreet} JD} (\bibinfo{year}{2009}), \bibinfo{month}{Sep.}
\bibinfo{title}{{Magnetic Fields of Nondegenerate Stars}}.
\bibinfo{journal}{{\em \araa}} \bibinfo{volume}{47} (\bibinfo{number}{1}): \bibinfo{pages}{333--370}. \bibinfo{doi}{\doi{10.1146/annurev-astro-082708-101833}}.
\eprint{0904.1938}.

\bibtype{Article}%
\bibitem[{Drout} et al.(2023)]{Drout+2023}
\bibinfo{author}{{Drout} MR}, \bibinfo{author}{{G{\"o}tberg} Y}, \bibinfo{author}{{Ludwig} BA}, \bibinfo{author}{{Groh} JH}, \bibinfo{author}{{de Mink} SE}, \bibinfo{author}{{O'Grady} AJG} and  \bibinfo{author}{{Smith} N} (\bibinfo{year}{2023}), \bibinfo{month}{Dec.}
\bibinfo{title}{{An observed population of intermediate-mass helium stars that have been stripped in binaries}}.
\bibinfo{journal}{{\em Science}} \bibinfo{volume}{382} (\bibinfo{number}{6676}): \bibinfo{pages}{1287--1291}. \bibinfo{doi}{\doi{10.1126/science.ade4970}}.
\eprint{2307.00061}.

\bibtype{Article}%
\bibitem[{Eggleton}(1983)]{Eggleton83}
\bibinfo{author}{{Eggleton} PP} (\bibinfo{year}{1983}), \bibinfo{month}{May}.
\bibinfo{title}{{Aproximations to the radii of Roche lobes.}}
\bibinfo{journal}{{\em \apj}} \bibinfo{volume}{268}: \bibinfo{pages}{368--369}. \bibinfo{doi}{\doi{10.1086/160960}}.

\bibtype{Article}%
\bibitem[{Fabry} et al.(2022)]{Fabry+22}
\bibinfo{author}{{Fabry} M}, \bibinfo{author}{{Marchant} P} and  \bibinfo{author}{{Sana} H} (\bibinfo{year}{2022}), \bibinfo{month}{May}.
\bibinfo{title}{{Modeling overcontact binaries. I. The effect of tidal deformation}}.
\bibinfo{journal}{{\em \aap}} \bibinfo{volume}{661}, \bibinfo{eid}{A123}. \bibinfo{doi}{\doi{10.1051/0004-6361/202243094}}.
\eprint{2202.08852}.

\bibtype{Article}%
\bibitem[{Farmer} et al.(2023)]{Farmer+2023}
\bibinfo{author}{{Farmer} R}, \bibinfo{author}{{Renzo} M}, \bibinfo{author}{{G{\"o}tberg} Y}, \bibinfo{author}{{Bellinger} E}, \bibinfo{author}{{Justham} S} and  \bibinfo{author}{{de Mink} SE} (\bibinfo{year}{2023}), \bibinfo{month}{Sep.}
\bibinfo{title}{{Observational predictions for Thorne-{\.Z}ytkow objects}}.
\bibinfo{journal}{{\em \mnras}} \bibinfo{volume}{524} (\bibinfo{number}{2}): \bibinfo{pages}{1692--1709}. \bibinfo{doi}{\doi{10.1093/mnras/stad1977}}.
\eprint{2305.07337}.

\bibtype{Article}%
\bibitem[{Ferrario} et al.(2009)]{Ferrario+2009}
\bibinfo{author}{{Ferrario} L}, \bibinfo{author}{{Pringle} JE}, \bibinfo{author}{{Tout} CA} and  \bibinfo{author}{{Wickramasinghe} DT} (\bibinfo{year}{2009}), \bibinfo{month}{Nov.}
\bibinfo{title}{{The origin of magnetism on the upper main sequence}}.
\bibinfo{journal}{{\em \mnras}} \bibinfo{volume}{400} (\bibinfo{number}{1}): \bibinfo{pages}{L71--L74}. \bibinfo{doi}{\doi{10.1111/j.1745-3933.2009.00765.x}}.

\bibtype{Article}%
\bibitem[{Frost} et al.(2024)]{Frost+2024}
\bibinfo{author}{{Frost} AJ}, \bibinfo{author}{{Sana} H}, \bibinfo{author}{{Mahy} L}, \bibinfo{author}{{Wade} G}, \bibinfo{author}{{Barron} J}, \bibinfo{author}{{Le Bouquin} JB}, \bibinfo{author}{{M{\'e}rand} A}, \bibinfo{author}{{Schneider} FRN}, \bibinfo{author}{{Shenar} T}, \bibinfo{author}{{Barb{\'a}} RH}, \bibinfo{author}{{Bowman} DM}, \bibinfo{author}{{Fabry} M}, \bibinfo{author}{{Farhang} A}, \bibinfo{author}{{Marchant} P}, \bibinfo{author}{{Morrell} NI} and  \bibinfo{author}{{Smoker} JV} (\bibinfo{year}{2024}), \bibinfo{month}{Apr.}
\bibinfo{title}{{A magnetic massive star has experienced a stellar merger}}.
\bibinfo{journal}{{\em Science}} \bibinfo{volume}{384} (\bibinfo{number}{6692}): \bibinfo{pages}{214--217}. \bibinfo{doi}{\doi{10.1126/science.adg7700}}.
\eprint{2404.10167}.

\bibtype{Article}%
\bibitem[{Fuller} and {Lai}(2012)]{FullerLai2012}
\bibinfo{author}{{Fuller} J} and  \bibinfo{author}{{Lai} D} (\bibinfo{year}{2012}), \bibinfo{month}{Mar.}
\bibinfo{title}{{Dynamical tides in eccentric binaries and tidally excited stellar pulsations in Kepler KOI-54}}.
\bibinfo{journal}{{\em \mnras}} \bibinfo{volume}{420} (\bibinfo{number}{4}): \bibinfo{pages}{3126--3138}. \bibinfo{doi}{\doi{10.1111/j.1365-2966.2011.20237.x}}.
\eprint{1107.4594}.

\bibtype{Article}%
\bibitem[{Gallegos-Garcia} et al.(2021)]{Gallegos-Garcia2021}
\bibinfo{author}{{Gallegos-Garcia} M}, \bibinfo{author}{{Berry} CPL}, \bibinfo{author}{{Marchant} P} and  \bibinfo{author}{{Kalogera} V} (\bibinfo{year}{2021}), \bibinfo{month}{Dec.}
\bibinfo{title}{{Binary Black Hole Formation with Detailed Modeling: Stable Mass Transfer Leads to Lower Merger Rates}}.
\bibinfo{journal}{{\em \apj}} \bibinfo{volume}{922} (\bibinfo{number}{2}), \bibinfo{eid}{110}. \bibinfo{doi}{\doi{10.3847/1538-4357/ac2610}}.
\eprint{2107.05702}.

\bibtype{Article}%
\bibitem[{Gallegos-Garcia} et al.(2023)]{Gallegos-Garcia+2023}
\bibinfo{author}{{Gallegos-Garcia} M}, \bibinfo{author}{{Berry} CPL} and  \bibinfo{author}{{Kalogera} V} (\bibinfo{year}{2023}), \bibinfo{month}{Oct.}
\bibinfo{title}{{Evolutionary Origins of Binary Neutron Star Mergers: Effects of Common Envelope Efficiency and Metallicity}}.
\bibinfo{journal}{{\em \apj}} \bibinfo{volume}{955} (\bibinfo{number}{2}), \bibinfo{eid}{133}. \bibinfo{doi}{\doi{10.3847/1538-4357/ace434}}.
\eprint{2211.15693}.

\bibtype{Article}%
\bibitem[{Garc{\'\i}a-Berro} et al.(2012)]{GarciaBerro+2012}
\bibinfo{author}{{Garc{\'\i}a-Berro} E}, \bibinfo{author}{{Lor{\'e}n-Aguilar} P}, \bibinfo{author}{{Aznar-Sigu{\'a}n} G}, \bibinfo{author}{{Torres} S}, \bibinfo{author}{{Camacho} J}, \bibinfo{author}{{Althaus} LG}, \bibinfo{author}{{C{\'o}rsico} AH}, \bibinfo{author}{{K{\"u}lebi} B} and  \bibinfo{author}{{Isern} J} (\bibinfo{year}{2012}), \bibinfo{month}{Apr.}
\bibinfo{title}{{Double Degenerate Mergers as Progenitors of High-field Magnetic White Dwarfs}}.
\bibinfo{journal}{{\em \apj}} \bibinfo{volume}{749} (\bibinfo{number}{1}), \bibinfo{eid}{25}. \bibinfo{doi}{\doi{10.1088/0004-637X/749/1/25}}.
\eprint{1202.0461}.

\bibtype{Article}%
\bibitem[{G{\"o}tberg} et al.(2020)]{Gotberg+2020}
\bibinfo{author}{{G{\"o}tberg} Y}, \bibinfo{author}{{de Mink} SE}, \bibinfo{author}{{McQuinn} M}, \bibinfo{author}{{Zapartas} E}, \bibinfo{author}{{Groh} JH} and  \bibinfo{author}{{Norman} C} (\bibinfo{year}{2020}), \bibinfo{month}{Feb.}
\bibinfo{title}{{Contribution from stars stripped in binaries to cosmic reionization of hydrogen and helium}}.
\bibinfo{journal}{{\em \aap}} \bibinfo{volume}{634}, \bibinfo{eid}{A134}. \bibinfo{doi}{\doi{10.1051/0004-6361/201936669}}.
\eprint{1911.00543}.

\bibtype{Article}%
\bibitem[{Hall} and {Evans}(2019)]{HallEvans2019}
\bibinfo{author}{{Hall} ED} and  \bibinfo{author}{{Evans} M} (\bibinfo{year}{2019}), \bibinfo{month}{Nov.}
\bibinfo{title}{{Metrics for next-generation gravitational-wave detectors}}.
\bibinfo{journal}{{\em Classical and Quantum Gravity}} \bibinfo{volume}{36} (\bibinfo{number}{22}), \bibinfo{eid}{225002}. \bibinfo{doi}{\doi{10.1088/1361-6382/ab41d6}}.
\eprint{1902.09485}.

\bibtype{Article}%
\bibitem[{Han} et al.(2002)]{Han+2002}
\bibinfo{author}{{Han} Z}, \bibinfo{author}{{Podsiadlowski} P}, \bibinfo{author}{{Maxted} PFL}, \bibinfo{author}{{Marsh} TR} and  \bibinfo{author}{{Ivanova} N} (\bibinfo{year}{2002}), \bibinfo{month}{Oct.}
\bibinfo{title}{{The origin of subdwarf B stars - I. The formation channels}}.
\bibinfo{journal}{{\em \mnras}} \bibinfo{volume}{336} (\bibinfo{number}{2}): \bibinfo{pages}{449--466}. \bibinfo{doi}{\doi{10.1046/j.1365-8711.2002.05752.x}}.
\eprint{astro-ph/0206130}.

\bibtype{Article}%
\bibitem[{Han} et al.(2007)]{Han+2007UV}
\bibinfo{author}{{Han} Z}, \bibinfo{author}{{Podsiadlowski} P} and  \bibinfo{author}{{Lynas-Gray} AE} (\bibinfo{year}{2007}), \bibinfo{month}{Sep.}
\bibinfo{title}{{A binary model for the UV-upturn of elliptical galaxies}}.
\bibinfo{journal}{{\em \mnras}} \bibinfo{volume}{380} (\bibinfo{number}{3}): \bibinfo{pages}{1098--1118}. \bibinfo{doi}{\doi{10.1111/j.1365-2966.2007.12151.x}}.
\eprint{0704.0863}.

\bibtype{Article}%
\bibitem[{Hopkins} et al.(2012)]{Hopkins+2012}
\bibinfo{author}{{Hopkins} PF}, \bibinfo{author}{{Quataert} E} and  \bibinfo{author}{{Murray} N} (\bibinfo{year}{2012}), \bibinfo{month}{Apr.}
\bibinfo{title}{{Stellar feedback in galaxies and the origin of galaxy-scale winds}}.
\bibinfo{journal}{{\em \mnras}} \bibinfo{volume}{421} (\bibinfo{number}{4}): \bibinfo{pages}{3522--3537}. \bibinfo{doi}{\doi{10.1111/j.1365-2966.2012.20593.x}}.
\eprint{1110.4638}.

\bibtype{Article}%
\bibitem[{Hoyle} and {Fowler}(1960)]{HoyleFowler1960}
\bibinfo{author}{{Hoyle} F} and  \bibinfo{author}{{Fowler} WA} (\bibinfo{year}{1960}), \bibinfo{month}{Nov.}
\bibinfo{title}{{Nucleosynthesis in Supernovae.}}
\bibinfo{journal}{{\em \apj}} \bibinfo{volume}{132}: \bibinfo{pages}{565}. \bibinfo{doi}{\doi{10.1086/146963}}.

\bibtype{Article}%
\bibitem[{Huang}(1966)]{Huang1966}
\bibinfo{author}{{Huang} SS} (\bibinfo{year}{1966}), \bibinfo{month}{Feb.}
\bibinfo{title}{{A theory of the origin and evolution of contact binaries}}.
\bibinfo{journal}{{\em Annales d'Astrophysique}} \bibinfo{volume}{29}: \bibinfo{pages}{331}.

\bibtype{Article}%
\bibitem[{Hurley} et al.(2002)]{Hurley+2002}
\bibinfo{author}{{Hurley} JR}, \bibinfo{author}{{Tout} CA} and  \bibinfo{author}{{Pols} OR} (\bibinfo{year}{2002}), \bibinfo{month}{Feb.}
\bibinfo{title}{{Evolution of binary stars and the effect of tides on binary populations}}.
\bibinfo{journal}{{\em \mnras}} \bibinfo{volume}{329} (\bibinfo{number}{4}): \bibinfo{pages}{897--928}. \bibinfo{doi}{\doi{10.1046/j.1365-8711.2002.05038.x}}.
\eprint{astro-ph/0201220}.

\bibtype{Article}%
\bibitem[{Iben} and {Tutukov}(1984)]{IbenTutukov1984}
\bibinfo{author}{{Iben} I. J} and  \bibinfo{author}{{Tutukov} AV} (\bibinfo{year}{1984}), \bibinfo{month}{Feb.}
\bibinfo{title}{{Supernovae of type I as end products of the evolution of binaries with components of moderate initial mass.}}
\bibinfo{journal}{{\em \apjs}} \bibinfo{volume}{54}: \bibinfo{pages}{335--372}. \bibinfo{doi}{\doi{10.1086/190932}}.

\bibtype{Article}%
\bibitem[{Istrate} et al.(2016)]{Istrate+2016}
\bibinfo{author}{{Istrate} AG}, \bibinfo{author}{{Marchant} P}, \bibinfo{author}{{Tauris} TM}, \bibinfo{author}{{Langer} N}, \bibinfo{author}{{Stancliffe} RJ} and  \bibinfo{author}{{Grassitelli} L} (\bibinfo{year}{2016}), \bibinfo{month}{Oct.}
\bibinfo{title}{{Models of low-mass helium white dwarfs including gravitational settling, thermal and chemical diffusion, and rotational mixing}}.
\bibinfo{journal}{{\em \aap}} \bibinfo{volume}{595}, \bibinfo{eid}{A35}. \bibinfo{doi}{\doi{10.1051/0004-6361/201628874}}.
\eprint{1606.04947}.

\bibtype{Article}%
\bibitem[{Ivanova} et al.(2013)]{Ivanova+2013}
\bibinfo{author}{{Ivanova} N}, \bibinfo{author}{{Justham} S}, \bibinfo{author}{{Chen} X}, \bibinfo{author}{{De Marco} O}, \bibinfo{author}{{Fryer} CL}, \bibinfo{author}{{Gaburov} E}, \bibinfo{author}{{Ge} H}, \bibinfo{author}{{Glebbeek} E}, \bibinfo{author}{{Han} Z}, \bibinfo{author}{{Li} XD}, \bibinfo{author}{{Lu} G}, \bibinfo{author}{{Marsh} T}, \bibinfo{author}{{Podsiadlowski} P}, \bibinfo{author}{{Potter} A}, \bibinfo{author}{{Soker} N}, \bibinfo{author}{{Taam} R}, \bibinfo{author}{{Tauris} TM}, \bibinfo{author}{{van den Heuvel} EPJ} and  \bibinfo{author}{{Webbink} RF} (\bibinfo{year}{2013}), \bibinfo{month}{Feb.}
\bibinfo{title}{{Common envelope evolution: where we stand and how we can move forward}}.
\bibinfo{journal}{{\em \aapr}} \bibinfo{volume}{21}, \bibinfo{eid}{59}. \bibinfo{doi}{\doi{10.1007/s00159-013-0059-2}}.
\eprint{1209.4302}.

\bibtype{Article}%
\bibitem[{Ivezi{\'c}} et al.(2019)]{Rubin2019}
\bibinfo{author}{{Ivezi{\'c}} {\v{Z}}}, \bibinfo{author}{{Kahn} SM}, \bibinfo{author}{{Tyson} JA} and  \bibinfo{author}{et. al.} (\bibinfo{year}{2019}), \bibinfo{month}{Mar.}
\bibinfo{title}{{LSST: From Science Drivers to Reference Design and Anticipated Data Products}}.
\bibinfo{journal}{{\em \apj}} \bibinfo{volume}{873} (\bibinfo{number}{2}), \bibinfo{eid}{111}. \bibinfo{doi}{\doi{10.3847/1538-4357/ab042c}}.
\eprint{0805.2366}.

\bibtype{Article}%
\bibitem[{Justham} et al.(2014)]{Justham+2014}
\bibinfo{author}{{Justham} S}, \bibinfo{author}{{Podsiadlowski} P} and  \bibinfo{author}{{Vink} JS} (\bibinfo{year}{2014}), \bibinfo{month}{Dec.}
\bibinfo{title}{{Luminous Blue Variables and Superluminous Supernovae from Binary Mergers}}.
\bibinfo{journal}{{\em \apj}} \bibinfo{volume}{796} (\bibinfo{number}{2}), \bibinfo{eid}{121}. \bibinfo{doi}{\doi{10.1088/0004-637X/796/2/121}}.
\eprint{1410.2426}.

\bibtype{Article}%
\bibitem[{Kippenhahn} and {Weigert}(1967)]{KippenhahnWeigert1967}
\bibinfo{author}{{Kippenhahn} R} and  \bibinfo{author}{{Weigert} A} (\bibinfo{year}{1967}), \bibinfo{month}{Jan.}
\bibinfo{title}{{Entwicklung in engen Doppelsternsystemen I. Massenaustausch vor und nach Beendigung des zentralen Wasserstoff-Brennens}}.
\bibinfo{journal}{{\em \zap}} \bibinfo{volume}{65}: \bibinfo{pages}{251}.

\bibtype{Article}%
\bibitem[{Klencki} et al.(2021)]{Klencki+2021}
\bibinfo{author}{{Klencki} J}, \bibinfo{author}{{Nelemans} G}, \bibinfo{author}{{Istrate} AG} and  \bibinfo{author}{{Chruslinska} M} (\bibinfo{year}{2021}), \bibinfo{month}{Jan.}
\bibinfo{title}{{It has to be cool: Supergiant progenitors of binary black hole mergers from common-envelope evolution}}.
\bibinfo{journal}{{\em \aap}} \bibinfo{volume}{645}, \bibinfo{eid}{A54}. \bibinfo{doi}{\doi{10.1051/0004-6361/202038707}}.
\eprint{2006.11286}.

\bibtype{Article}%
\bibitem[{Kuiper}(1941)]{Kuiper1941}
\bibinfo{author}{{Kuiper} GP} (\bibinfo{year}{1941}), \bibinfo{month}{Jan.}
\bibinfo{title}{{On the Interpretation of {\ensuremath{\beta}} Lyrae and Other Close Binaries.}}
\bibinfo{journal}{{\em \apj}} \bibinfo{volume}{93}: \bibinfo{pages}{133}. \bibinfo{doi}{\doi{10.1086/144252}}.

\bibtype{Article}%
\bibitem[{Kulkarni} et al.(2021)]{Kulkarni+2021}
\bibinfo{author}{{Kulkarni} SR}, \bibinfo{author}{{Harrison} FA}, \bibinfo{author}{{Grefenstette} BW}, \bibinfo{author}{{Earnshaw} HP}, \bibinfo{author}{{Andreoni} I}, \bibinfo{author}{{Berg} DA}, \bibinfo{author}{{Bloom} JS}, \bibinfo{author}{{Cenko} SB}, \bibinfo{author}{{Chornock} R}, \bibinfo{author}{{Christiansen} JL}, \bibinfo{author}{{Coughlin} MW}, \bibinfo{author}{{Wuollet Criswell} A}, \bibinfo{author}{{Darvish} B}, \bibinfo{author}{{Das} KK}, \bibinfo{author}{{De} K}, \bibinfo{author}{{Dessart} L}, \bibinfo{author}{{Dixon} D}, \bibinfo{author}{{Dorsman} B}, \bibinfo{author}{{El-Badry} K}, \bibinfo{author}{{Evans} C}, \bibinfo{author}{{Ford} KES}, \bibinfo{author}{{Fremling} C}, \bibinfo{author}{{Gansicke} BT}, \bibinfo{author}{{Gezari} S}, \bibinfo{author}{{Goetberg} Y}, \bibinfo{author}{{Green} GM}, \bibinfo{author}{{Graham} MJ}, \bibinfo{author}{{Heida} M}, \bibinfo{author}{{Ho} AYQ}, \bibinfo{author}{{Jaodand} AD}, \bibinfo{author}{{Johns-Krull} CM}, \bibinfo{author}{{Kasliwal} MM},
  \bibinfo{author}{{Lazzarini} M}, \bibinfo{author}{{Lu} W}, \bibinfo{author}{{Margutti} R}, \bibinfo{author}{{Martin} DC}, \bibinfo{author}{{Masters} DC}, \bibinfo{author}{{McKernan} B}, \bibinfo{author}{{Naze} Y}, \bibinfo{author}{{Nissanke} SM}, \bibinfo{author}{{Parazin} B}, \bibinfo{author}{{Perley} DA}, \bibinfo{author}{{Phinney} ES}, \bibinfo{author}{{Piro} AL}, \bibinfo{author}{{Raaijmakers} G}, \bibinfo{author}{{Rauw} G}, \bibinfo{author}{{Rodriguez} AC}, \bibinfo{author}{{Sana} H}, \bibinfo{author}{{Senchyna} P}, \bibinfo{author}{{Singer} LP}, \bibinfo{author}{{Spake} JJ}, \bibinfo{author}{{Stassun} KG}, \bibinfo{author}{{Stern} D}, \bibinfo{author}{{Teplitz} HI}, \bibinfo{author}{{Weisz} DR} and  \bibinfo{author}{{Yao} Y} (\bibinfo{year}{2021}), \bibinfo{month}{Nov.}
\bibinfo{title}{{Science with the Ultraviolet Explorer (UVEX)}}.
\bibinfo{journal}{{\em arXiv e-prints}} , \bibinfo{eid}{arXiv:2111.15608}\bibinfo{doi}{\doi{10.48550/arXiv.2111.15608}}.
\eprint{2111.15608}.

\bibtype{Article}%
\bibitem[{Kummer} et al.(2023)]{Kummer+2023}
\bibinfo{author}{{Kummer} F}, \bibinfo{author}{{Toonen} S} and  \bibinfo{author}{{de Koter} A} (\bibinfo{year}{2023}), \bibinfo{month}{Oct.}
\bibinfo{title}{{The main evolutionary pathways of massive hierarchical triple stars}}.
\bibinfo{journal}{{\em \aap}} \bibinfo{volume}{678}, \bibinfo{eid}{A60}. \bibinfo{doi}{\doi{10.1051/0004-6361/202347179}}.
\eprint{2306.09400}.

\bibtype{Article}%
\bibitem[{Kupfer} et al.(2018)]{Kupfer+2018}
\bibinfo{author}{{Kupfer} T}, \bibinfo{author}{{Korol} V}, \bibinfo{author}{{Shah} S}, \bibinfo{author}{{Nelemans} G}, \bibinfo{author}{{Marsh} TR}, \bibinfo{author}{{Ramsay} G}, \bibinfo{author}{{Groot} PJ}, \bibinfo{author}{{Steeghs} DTH} and  \bibinfo{author}{{Rossi} EM} (\bibinfo{year}{2018}), \bibinfo{month}{Oct.}
\bibinfo{title}{{LISA verification binaries with updated distances from Gaia Data Release 2}}.
\bibinfo{journal}{{\em \mnras}} \bibinfo{volume}{480} (\bibinfo{number}{1}): \bibinfo{pages}{302--309}. \bibinfo{doi}{\doi{10.1093/mnras/sty1545}}.
\eprint{1805.00482}.

\bibtype{Inproceedings}%
\bibitem[{Langer}(1997)]{Langer1997}
\bibinfo{author}{{Langer} N} (\bibinfo{year}{1997}), \bibinfo{month}{Jan.}, \bibinfo{title}{{The Eddington Limit in Rotating Massive Stars}}, \bibinfo{editor}{{Nota} A} and  \bibinfo{editor}{{Lamers} H}, (Eds.), \bibinfo{booktitle}{Luminous Blue Variables: Massive Stars in Transition}, \bibinfo{series}{Astronomical Society of the Pacific Conference Series}, \bibinfo{volume}{120}, pp.~\bibinfo{pages}{83}.

\bibtype{Article}%
\bibitem[{Langer}(2012)]{Langer2012}
\bibinfo{author}{{Langer} N} (\bibinfo{year}{2012}), \bibinfo{month}{Sep.}
\bibinfo{title}{{Presupernova Evolution of Massive Single and Binary Stars}}.
\bibinfo{journal}{{\em \araa}} \bibinfo{volume}{50}: \bibinfo{pages}{107--164}. \bibinfo{doi}{\doi{10.1146/annurev-astro-081811-125534}}.
\eprint{1206.5443}.

\bibtype{Article}%
\bibitem[{Lauterborn}(1970)]{Lauterborn1970}
\bibinfo{author}{{Lauterborn} D} (\bibinfo{year}{1970}), \bibinfo{month}{Jul.}
\bibinfo{title}{{Evolution with mass exchange of case C for a binary system of total mass 7 M sun.}}
\bibinfo{journal}{{\em \aap}} \bibinfo{volume}{7}: \bibinfo{pages}{150}.

\bibtype{Article}%
\bibitem[{Leiner} et al.(2019)]{Leiner+2019}
\bibinfo{author}{{Leiner} E}, \bibinfo{author}{{Mathieu} RD}, \bibinfo{author}{{Vanderburg} A}, \bibinfo{author}{{Gosnell} NM} and  \bibinfo{author}{{Smith} JC} (\bibinfo{year}{2019}), \bibinfo{month}{Aug.}
\bibinfo{title}{{Blue Lurkers: Hidden Blue Stragglers on the M67 Main Sequence Identified from Their Kepler/K2 Rotation Periods}}.
\bibinfo{journal}{{\em \apj}} \bibinfo{volume}{881} (\bibinfo{number}{1}), \bibinfo{eid}{47}. \bibinfo{doi}{\doi{10.3847/1538-4357/ab2bf8}}.
\eprint{1904.02169}.

\bibtype{Article}%
\bibitem[{Li} et al.(2019)]{Li+2019}
\bibinfo{author}{{Li} Z}, \bibinfo{author}{{Chen} X}, \bibinfo{author}{{Chen} HL} and  \bibinfo{author}{{Han} Z} (\bibinfo{year}{2019}), \bibinfo{month}{Feb.}
\bibinfo{title}{{Formation of Extremely Low-mass White Dwarfs in Double Degenerates}}.
\bibinfo{journal}{{\em \apj}} \bibinfo{volume}{871} (\bibinfo{number}{2}), \bibinfo{eid}{148}. \bibinfo{doi}{\doi{10.3847/1538-4357/aaf9a1}}.
\eprint{1812.07226}.

\bibtype{Article}%
\bibitem[{Liu} et al.(2023)]{Liu+2023}
\bibinfo{author}{{Liu} ZW}, \bibinfo{author}{{R{\"o}pke} FK} and  \bibinfo{author}{{Han} Z} (\bibinfo{year}{2023}), \bibinfo{month}{Aug.}
\bibinfo{title}{{Type Ia Supernova Explosions in Binary Systems: A Review}}.
\bibinfo{journal}{{\em Research in Astronomy and Astrophysics}} \bibinfo{volume}{23} (\bibinfo{number}{8}), \bibinfo{eid}{082001}. \bibinfo{doi}{\doi{10.1088/1674-4527/acd89e}}.
\eprint{2305.13305}.

\bibtype{Article}%
\bibitem[{Lubow} and {Shu}(1975)]{LubowShu1975}
\bibinfo{author}{{Lubow} SH} and  \bibinfo{author}{{Shu} FH} (\bibinfo{year}{1975}), \bibinfo{month}{Jun.}
\bibinfo{title}{{Gas dynamics of semidetached binaries.}}
\bibinfo{journal}{{\em \apj}} \bibinfo{volume}{198}: \bibinfo{pages}{383--405}. \bibinfo{doi}{\doi{10.1086/153614}}.

\bibtype{Article}%
\bibitem[{Ma} et al.(2024)]{JingZe+2024}
\bibinfo{author}{{Ma} JZ}, \bibinfo{author}{{Chiavassa} A}, \bibinfo{author}{{de Mink} SE}, \bibinfo{author}{{Valli} R}, \bibinfo{author}{{Justham} S} and  \bibinfo{author}{{Freytag} B} (\bibinfo{year}{2024}), \bibinfo{month}{Feb.}
\bibinfo{title}{{Is Betelgeuse Really Rotating? Synthetic ALMA Observations of Large-scale Convection in 3D Simulations of Red Supergiants}}.
\bibinfo{journal}{{\em \apjl}} \bibinfo{volume}{962} (\bibinfo{number}{2}), \bibinfo{eid}{L36}. \bibinfo{doi}{\doi{10.3847/2041-8213/ad24fd}}.
\eprint{2311.16885}.

\bibtype{Article}%
\bibitem[{Maeder} et al.(2014)]{Maeder+2014}
\bibinfo{author}{{Maeder} A}, \bibinfo{author}{{Przybilla} N}, \bibinfo{author}{{Nieva} MF}, \bibinfo{author}{{Georgy} C}, \bibinfo{author}{{Meynet} G}, \bibinfo{author}{{Ekstr{\"o}m} S} and  \bibinfo{author}{{Eggenberger} P} (\bibinfo{year}{2014}), \bibinfo{month}{May}.
\bibinfo{title}{{Evolution of surface CNO abundances in massive stars}}.
\bibinfo{journal}{{\em \aap}} \bibinfo{volume}{565}, \bibinfo{eid}{A39}. \bibinfo{doi}{\doi{10.1051/0004-6361/201220602}}.
\eprint{1404.1020}.

\bibtype{Article}%
\bibitem[{Maiolino} and {Mannucci}(2019)]{MaiolinoMannucci2019}
\bibinfo{author}{{Maiolino} R} and  \bibinfo{author}{{Mannucci} F} (\bibinfo{year}{2019}), \bibinfo{month}{Feb.}
\bibinfo{title}{{De re metallica: the cosmic chemical evolution of galaxies}}.
\bibinfo{journal}{{\em \aapr}} \bibinfo{volume}{27} (\bibinfo{number}{1}), \bibinfo{eid}{3}. \bibinfo{doi}{\doi{10.1007/s00159-018-0112-2}}.
\eprint{1811.09642}.

\bibtype{Article}%
\bibitem[{Mandel} and {de Mink}(2016)]{MandelDeMink2016}
\bibinfo{author}{{Mandel} I} and  \bibinfo{author}{{de Mink} SE} (\bibinfo{year}{2016}), \bibinfo{month}{May}.
\bibinfo{title}{{Merging binary black holes formed through chemically homogeneous evolution in short-period stellar binaries}}.
\bibinfo{journal}{{\em \mnras}} \bibinfo{volume}{458} (\bibinfo{number}{3}): \bibinfo{pages}{2634--2647}. \bibinfo{doi}{\doi{10.1093/mnras/stw379}}.
\eprint{1601.00007}.

\bibtype{Article}%
\bibitem[{Mandel} and {Farmer}(2022)]{MandelFarmer2022}
\bibinfo{author}{{Mandel} I} and  \bibinfo{author}{{Farmer} A} (\bibinfo{year}{2022}), \bibinfo{month}{Apr.}
\bibinfo{title}{{Merging stellar-mass binary black holes}}.
\bibinfo{journal}{{\em \physrep}} \bibinfo{volume}{955}: \bibinfo{pages}{1--24}. \bibinfo{doi}{\doi{10.1016/j.physrep.2022.01.003}}.
\eprint{1806.05820}.

\bibtype{Article}%
\bibitem[{Maoz} et al.(2014)]{Maoz+2014}
\bibinfo{author}{{Maoz} D}, \bibinfo{author}{{Mannucci} F} and  \bibinfo{author}{{Nelemans} G} (\bibinfo{year}{2014}), \bibinfo{month}{Aug.}
\bibinfo{title}{{Observational Clues to the Progenitors of Type Ia Supernovae}}.
\bibinfo{journal}{{\em \araa}} \bibinfo{volume}{52}: \bibinfo{pages}{107--170}. \bibinfo{doi}{\doi{10.1146/annurev-astro-082812-141031}}.
\eprint{1312.0628}.

\bibtype{Article}%
\bibitem[{Marchant} and {Bodensteiner}(2024)]{Marchant2024}
\bibinfo{author}{{Marchant} P} and  \bibinfo{author}{{Bodensteiner} J} (\bibinfo{year}{2024}), \bibinfo{month}{Sep.}
\bibinfo{title}{{The Evolution of Massive Binary Stars}}.
\bibinfo{journal}{{\em \araa}} \bibinfo{volume}{62} (\bibinfo{number}{1}): \bibinfo{pages}{21--61}. \bibinfo{doi}{\doi{10.1146/annurev-astro-052722-105936}}.
\eprint{2311.01865}.

\bibtype{Article}%
\bibitem[{Marchant} et al.(2016)]{Marchant+2016}
\bibinfo{author}{{Marchant} P}, \bibinfo{author}{{Langer} N}, \bibinfo{author}{{Podsiadlowski} P}, \bibinfo{author}{{Tauris} TM} and  \bibinfo{author}{{Moriya} TJ} (\bibinfo{year}{2016}), \bibinfo{month}{Apr.}
\bibinfo{title}{{A new route towards merging massive black holes}}.
\bibinfo{journal}{{\em \aap}} \bibinfo{volume}{588}, \bibinfo{eid}{A50}. \bibinfo{doi}{\doi{10.1051/0004-6361/201628133}}.
\eprint{1601.03718}.

\bibtype{Article}%
\bibitem[{Marchant} et al.(2017)]{Marchant+2017}
\bibinfo{author}{{Marchant} P}, \bibinfo{author}{{Langer} N}, \bibinfo{author}{{Podsiadlowski} P}, \bibinfo{author}{{Tauris} TM}, \bibinfo{author}{{de Mink} S}, \bibinfo{author}{{Mandel} I} and  \bibinfo{author}{{Moriya} TJ} (\bibinfo{year}{2017}), \bibinfo{month}{Aug.}
\bibinfo{title}{{Ultra-luminous X-ray sources and neutron-star-black-hole mergers from very massive close binaries at low metallicity}}.
\bibinfo{journal}{{\em \aap}} \bibinfo{volume}{604}, \bibinfo{eid}{A55}. \bibinfo{doi}{\doi{10.1051/0004-6361/201630188}}.
\eprint{1705.04734}.

\bibtype{Article}%
\bibitem[{Marchant} et al.(2021)]{Marchant+2021}
\bibinfo{author}{{Marchant} P}, \bibinfo{author}{{Pappas} KMW}, \bibinfo{author}{{Gallegos-Garcia} M}, \bibinfo{author}{{Berry} CPL}, \bibinfo{author}{{Taam} RE}, \bibinfo{author}{{Kalogera} V} and  \bibinfo{author}{{Podsiadlowski} P} (\bibinfo{year}{2021}), \bibinfo{month}{Jun.}
\bibinfo{title}{{The role of mass transfer and common envelope evolution in the formation of merging binary black holes}}.
\bibinfo{journal}{{\em \aap}} \bibinfo{volume}{650}, \bibinfo{eid}{A107}. \bibinfo{doi}{\doi{10.1051/0004-6361/202039992}}.
\eprint{2103.09243}.

\bibtype{Article}%
\bibitem[{Mateo} et al.(1990)]{Mateo+1990}
\bibinfo{author}{{Mateo} M}, \bibinfo{author}{{Harris} HC}, \bibinfo{author}{{Nemec} J} and  \bibinfo{author}{{Olszewski} EW} (\bibinfo{year}{1990}), \bibinfo{month}{Aug.}
\bibinfo{title}{{Blue Stragglers as Remnants of Stellar Mergers: The Discovery of Short-Period Eclipsing Binaries in the Globular Cluster NGC 5466}}.
\bibinfo{journal}{{\em \aj}} \bibinfo{volume}{100}: \bibinfo{pages}{469}. \bibinfo{doi}{\doi{10.1086/115530}}.

\bibtype{Article}%
\bibitem[{Menon} and {Heger}(2017)]{MenonHeger2017}
\bibinfo{author}{{Menon} A} and  \bibinfo{author}{{Heger} A} (\bibinfo{year}{2017}), \bibinfo{month}{Jan.}
\bibinfo{title}{{The quest for blue supergiants: binary merger models for the evolution of the progenitor of SN 1987A.}}
\bibinfo{journal}{{\em \mnras}} \bibinfo{volume}{469}: \bibinfo{pages}{4649--4664}. \bibinfo{doi}{\doi{10.1093/mnras/stx818}}.
\eprint{1703.04918}.

\bibtype{Article}%
\bibitem[{Miller-Jones} et al.(2021)]{Miller-Jones2021}
\bibinfo{author}{{Miller-Jones} JCA}, \bibinfo{author}{{Bahramian} A}, \bibinfo{author}{{Orosz} JA}, \bibinfo{author}{{Mandel} I}, \bibinfo{author}{{Gou} L}, \bibinfo{author}{{Maccarone} TJ}, \bibinfo{author}{{Neijssel} CJ}, \bibinfo{author}{{Zhao} X}, \bibinfo{author}{{Zi{\'o}{\l}kowski} J}, \bibinfo{author}{{Reid} MJ}, \bibinfo{author}{{Uttley} P}, \bibinfo{author}{{Zheng} X}, \bibinfo{author}{{Byun} DY}, \bibinfo{author}{{Dodson} R}, \bibinfo{author}{{Grinberg} V}, \bibinfo{author}{{Jung} T}, \bibinfo{author}{{Kim} JS}, \bibinfo{author}{{Marcote} B}, \bibinfo{author}{{Markoff} S}, \bibinfo{author}{{Rioja} MJ}, \bibinfo{author}{{Rushton} AP}, \bibinfo{author}{{Russell} DM}, \bibinfo{author}{{Sivakoff} GR}, \bibinfo{author}{{Tetarenko} AJ}, \bibinfo{author}{{Tudose} V} and  \bibinfo{author}{{Wilms} J} (\bibinfo{year}{2021}), \bibinfo{month}{Mar.}
\bibinfo{title}{{Cygnus X-1 contains a 21-solar mass black hole{\textemdash}Implications for massive star winds}}.
\bibinfo{journal}{{\em Science}} \bibinfo{volume}{371} (\bibinfo{number}{6533}): \bibinfo{pages}{1046--1049}. \bibinfo{doi}{\doi{10.1126/science.abb3363}}.
\eprint{2102.09091}.

\bibtype{Article}%
\bibitem[{Mourard} et al.(2018)]{Mourard+2018}
\bibinfo{author}{{Mourard} D}, \bibinfo{author}{{Bro{\v{z}}} M}, \bibinfo{author}{{Nemravov{\'a}} JA}, \bibinfo{author}{{Harmanec} P}, \bibinfo{author}{{Budaj} J}, \bibinfo{author}{{Baron} F}, \bibinfo{author}{{Monnier} JD}, \bibinfo{author}{{Schaefer} GH}, \bibinfo{author}{{Schmitt} H}, \bibinfo{author}{{Tallon-Bosc} I}, \bibinfo{author}{{Armstrong} JT}, \bibinfo{author}{{Baines} EK}, \bibinfo{author}{{Bonneau} D}, \bibinfo{author}{{Bo{\v{z}}i{\'c}} H}, \bibinfo{author}{{Clausse} JM}, \bibinfo{author}{{Farrington} C}, \bibinfo{author}{{Gies} D}, \bibinfo{author}{{Jury{\v{s}}ek} J}, \bibinfo{author}{{Kor{\v{c}}{\'a}kov{\'a}} D}, \bibinfo{author}{{McAlister} H}, \bibinfo{author}{{Meilland} A}, \bibinfo{author}{{Nardetto} N}, \bibinfo{author}{{Svoboda} P}, \bibinfo{author}{{{\v{S}}lechta} M}, \bibinfo{author}{{Wolf} M} and  \bibinfo{author}{{Zasche} P} (\bibinfo{year}{2018}), \bibinfo{month}{Oct.}
\bibinfo{title}{{Physical properties of {\ensuremath{\beta}} Lyrae A and its opaque accretion disk}}.
\bibinfo{journal}{{\em \aap}} \bibinfo{volume}{618}, \bibinfo{eid}{A112}. \bibinfo{doi}{\doi{10.1051/0004-6361/201832952}}.
\eprint{1807.04789}.

\bibtype{Article}%
\bibitem[{Packet}(1981)]{Packet1981}
\bibinfo{author}{{Packet} W} (\bibinfo{year}{1981}), \bibinfo{month}{Sep.}
\bibinfo{title}{{On the spin-up of the mass accreting component in a close binary system}}.
\bibinfo{journal}{{\em \aap}} \bibinfo{volume}{102} (\bibinfo{number}{1}): \bibinfo{pages}{17--19}.

\bibtype{Article}%
\bibitem[{Paczy{\'n}ski}(1967)]{Paczynski1967b}
\bibinfo{author}{{Paczy{\'n}ski} B} (\bibinfo{year}{1967}), \bibinfo{month}{Jan.}
\bibinfo{title}{{Evolution of Close Binaries. V. The Evolution of Massive Binaries and the Formation of the Wolf-Rayet Stars}}.
\bibinfo{journal}{{\em \actaa}} \bibinfo{volume}{17}: \bibinfo{pages}{355}.

\bibtype{Inproceedings}%
\bibitem[{Paczynski}(1976)]{Paczynski1976}
\bibinfo{author}{{Paczynski} B} (\bibinfo{year}{1976}), \bibinfo{month}{Jan.}, \bibinfo{title}{{Common Envelope Binaries}}, \bibinfo{editor}{{Eggleton} P}, \bibinfo{editor}{{Mitton} S} and  \bibinfo{editor}{{Whelan} J}, (Eds.), \bibinfo{booktitle}{Structure and Evolution of Close Binary Systems}, \bibinfo{series}{IAU Symposium}, \bibinfo{volume}{73}, pp.~\bibinfo{pages}{75}.

\bibtype{Article}%
\bibitem[{Paczynski}(1991)]{Paczynski1991}
\bibinfo{author}{{Paczynski} B} (\bibinfo{year}{1991}), \bibinfo{month}{Apr.}
\bibinfo{title}{{A Polytropic Model of an Accretion Disk, a Boundary Layer, and a Star}}.
\bibinfo{journal}{{\em \apj}} \bibinfo{volume}{370}: \bibinfo{pages}{597}. \bibinfo{doi}{\doi{10.1086/169846}}.

\bibtype{Article}%
\bibitem[{Panagia} et al.(1987)]{Panagia+1987}
\bibinfo{author}{{Panagia} N}, \bibinfo{author}{{Gilmozzi} R}, \bibinfo{author}{{Clavel} J}, \bibinfo{author}{{Barylak} M}, \bibinfo{author}{{Gonzalez-Riestra} R}, \bibinfo{author}{{Lloyd} C}, \bibinfo{author}{{Sanz Fernandez de Cordoba} L} and  \bibinfo{author}{{Wamsteker} W} (\bibinfo{year}{1987}), \bibinfo{month}{May}.
\bibinfo{title}{{Photometric properties of SN 1987A and other sources in the same field.}}
\bibinfo{journal}{{\em \aap}} \bibinfo{volume}{177}: \bibinfo{pages}{L25--L28}.

\bibtype{Article}%
\bibitem[{Panuzzo} et al.(2024)]{Panuzzo+2024}
\bibinfo{author}{{Panuzzo} P}, \bibinfo{author}{{Mazeh} T}, \bibinfo{author}{{Arenou}} and  \bibinfo{author}{et. al.} (\bibinfo{year}{2024}), \bibinfo{month}{Jun.}
\bibinfo{title}{{Discovery of a dormant 33 solar-mass black hole in pre-release Gaia astrometry}}.
\bibinfo{journal}{{\em \aap}} \bibinfo{volume}{686}, \bibinfo{eid}{L2}. \bibinfo{doi}{\doi{10.1051/0004-6361/202449763}}.
\eprint{2404.10486}.

\bibtype{Article}%
\bibitem[{Pastorello} et al.(2019)]{Pastorello+2019}
\bibinfo{author}{{Pastorello} A}, \bibinfo{author}{{Mason} E}, \bibinfo{author}{{Taubenberger} S}, \bibinfo{author}{{Fraser} M}, \bibinfo{author}{{Cortini} G}, \bibinfo{author}{{Tomasella} L}, \bibinfo{author}{{Botticella} MT}, \bibinfo{author}{{Elias-Rosa} N}, \bibinfo{author}{{Kotak} R}, \bibinfo{author}{{Smartt} SJ}, \bibinfo{author}{{Benetti} S}, \bibinfo{author}{{Cappellaro} E}, \bibinfo{author}{{Turatto} M}, \bibinfo{author}{{Tartaglia} L}, \bibinfo{author}{{Djorgovski} SG}, \bibinfo{author}{{Drake} AJ}, \bibinfo{author}{{Berton} M}, \bibinfo{author}{{Briganti} F}, \bibinfo{author}{{Brimacombe} J}, \bibinfo{author}{{Bufano} F}, \bibinfo{author}{{Cai} YZ}, \bibinfo{author}{{Chen} S}, \bibinfo{author}{{Christensen} EJ}, \bibinfo{author}{{Ciabattari} F}, \bibinfo{author}{{Congiu} E}, \bibinfo{author}{{Dimai} A}, \bibinfo{author}{{Inserra} C}, \bibinfo{author}{{Kankare} E}, \bibinfo{author}{{Magill} L}, \bibinfo{author}{{Maguire} K}, \bibinfo{author}{{Martinelli} F}, \bibinfo{author}{{Morales-Garoffolo} A},
  \bibinfo{author}{{Ochner} P}, \bibinfo{author}{{Pignata} G}, \bibinfo{author}{{Reguitti} A}, \bibinfo{author}{{Sollerman} J}, \bibinfo{author}{{Spiro} S}, \bibinfo{author}{{Terreran} G} and  \bibinfo{author}{{Wright} DE} (\bibinfo{year}{2019}), \bibinfo{month}{Oct.}
\bibinfo{title}{{Luminous red novae: Stellar mergers or giant eruptions?}}
\bibinfo{journal}{{\em \aap}} \bibinfo{volume}{630}, \bibinfo{eid}{A75}. \bibinfo{doi}{\doi{10.1051/0004-6361/201935999}}.
\eprint{1906.00812}.

\bibtype{Article}%
\bibitem[{Pauli} et al.(2022)]{Pauli+2022}
\bibinfo{author}{{Pauli} D}, \bibinfo{author}{{Langer} N}, \bibinfo{author}{{Aguilera-Dena} DR}, \bibinfo{author}{{Wang} C} and  \bibinfo{author}{{Marchant} P} (\bibinfo{year}{2022}), \bibinfo{month}{Nov.}
\bibinfo{title}{{A synthetic population of Wolf-Rayet stars in the LMC based on detailed single and binary star evolution models}}.
\bibinfo{journal}{{\em \aap}} \bibinfo{volume}{667}, \bibinfo{eid}{A58}. \bibinfo{doi}{\doi{10.1051/0004-6361/202243965}}.
\eprint{2208.10194}.

\bibtype{Article}%
\bibitem[{Pelisoli} et al.(2020)]{Pelisoli2020}
\bibinfo{author}{{Pelisoli} I}, \bibinfo{author}{{Vos} J}, \bibinfo{author}{{Geier} S}, \bibinfo{author}{{Schaffenroth} V} and  \bibinfo{author}{{Baran} AS} (\bibinfo{year}{2020}), \bibinfo{month}{Oct.}
\bibinfo{title}{{Alone but not lonely: Observational evidence that binary interaction is always required to form hot subdwarf stars}}.
\bibinfo{journal}{{\em \aap}} \bibinfo{volume}{642}, \bibinfo{eid}{A180}. \bibinfo{doi}{\doi{10.1051/0004-6361/202038473}}.
\eprint{2008.07522}.

\bibtype{Article}%
\bibitem[{Pelisoli} et al.(2022)]{Pelisoli+2022}
\bibinfo{author}{{Pelisoli} I}, \bibinfo{author}{{Dorsch} M}, \bibinfo{author}{{Heber} U}, \bibinfo{author}{{G{\"a}nsicke} B}, \bibinfo{author}{{Geier} S}, \bibinfo{author}{{Kupfer} T}, \bibinfo{author}{{N{\'e}meth} P}, \bibinfo{author}{{Scaringi} S} and  \bibinfo{author}{{Schaffenroth} V} (\bibinfo{year}{2022}), \bibinfo{month}{Sep.}
\bibinfo{title}{{Discovery and analysis of three magnetic hot subdwarf stars: evidence for merger-induced magnetic fields}}.
\bibinfo{journal}{{\em \mnras}} \bibinfo{volume}{515} (\bibinfo{number}{2}): \bibinfo{pages}{2496--2510}. \bibinfo{doi}{\doi{10.1093/mnras/stac1069}}.
\eprint{2204.06575}.

\bibtype{Article}%
\bibitem[{Peters}(1964)]{Peters1964}
\bibinfo{author}{{Peters} PC} (\bibinfo{year}{1964}), \bibinfo{month}{Nov.}
\bibinfo{title}{{Gravitational Radiation and the Motion of Two Point Masses}}.
\bibinfo{journal}{{\em Physical Review}} \bibinfo{volume}{136} (\bibinfo{number}{4B}): \bibinfo{pages}{1224--1232}. \bibinfo{doi}{\doi{10.1103/PhysRev.136.B1224}}.

\bibtype{Article}%
\bibitem[{Picco} et al.(2024)]{Picco+2024}
\bibinfo{author}{{Picco} A}, \bibinfo{author}{{Marchant} P}, \bibinfo{author}{{Sana} H} and  \bibinfo{author}{{Nelemans} G} (\bibinfo{year}{2024}), \bibinfo{month}{Jan.}
\bibinfo{title}{{Forming merging double compact objects with stable mass transfer}}.
\bibinfo{journal}{{\em \aap}} \bibinfo{volume}{681}, \bibinfo{eid}{A31}. \bibinfo{doi}{\doi{10.1051/0004-6361/202347090}}.
\eprint{2309.05736}.

\bibtype{Article}%
\bibitem[{Podsiadlowski} and {Joss}(1989)]{PodsiadlowskiJoss1989}
\bibinfo{author}{{Podsiadlowski} P} and  \bibinfo{author}{{Joss} PC} (\bibinfo{year}{1989}), \bibinfo{month}{Mar.}
\bibinfo{title}{{An alternative binary model for SN1987A}}.
\bibinfo{journal}{{\em \nat}} \bibinfo{volume}{338} (\bibinfo{number}{6214}): \bibinfo{pages}{401--403}. \bibinfo{doi}{\doi{10.1038/338401a0}}.

\bibtype{Article}%
\bibitem[{Podsiadlowski} et al.(1992)]{Podsiadlowski+1992}
\bibinfo{author}{{Podsiadlowski} P}, \bibinfo{author}{{Joss} PC} and  \bibinfo{author}{{Hsu} JJL} (\bibinfo{year}{1992}), \bibinfo{month}{May}.
\bibinfo{title}{{Presupernova Evolution in Massive Interacting Binaries}}.
\bibinfo{journal}{{\em \apj}} \bibinfo{volume}{391}: \bibinfo{pages}{246}. \bibinfo{doi}{\doi{10.1086/171341}}.

\bibtype{Article}%
\bibitem[{Popham} and {Narayan}(1991)]{PophamNarayan1991}
\bibinfo{author}{{Popham} R} and  \bibinfo{author}{{Narayan} R} (\bibinfo{year}{1991}), \bibinfo{month}{Apr.}
\bibinfo{title}{{Does Accretion Cease When a Star Approaches Breakup?}}
\bibinfo{journal}{{\em \apj}} \bibinfo{volume}{370}: \bibinfo{pages}{604}. \bibinfo{doi}{\doi{10.1086/169847}}.

\bibtype{Article}%
\bibitem[{Pr{\v{s}}a} and {Zwitter}(2005)]{PrsaZwitter2005}
\bibinfo{author}{{Pr{\v{s}}a} A} and  \bibinfo{author}{{Zwitter} T} (\bibinfo{year}{2005}), \bibinfo{month}{Jul.}
\bibinfo{title}{{A Computational Guide to Physics of Eclipsing Binaries. I. Demonstrations and Perspectives}}.
\bibinfo{journal}{{\em \apj}} \bibinfo{volume}{628} (\bibinfo{number}{1}): \bibinfo{pages}{426--438}. \bibinfo{doi}{\doi{10.1086/430591}}.
\eprint{astro-ph/0503361}.

\bibtype{Article}%
\bibitem[{Reig}(2011)]{Reig2011}
\bibinfo{author}{{Reig} P} (\bibinfo{year}{2011}), \bibinfo{month}{Mar.}
\bibinfo{title}{{Be/X-ray binaries}}.
\bibinfo{journal}{{\em \apss}} \bibinfo{volume}{332} (\bibinfo{number}{1}): \bibinfo{pages}{1--29}. \bibinfo{doi}{\doi{10.1007/s10509-010-0575-8}}.
\eprint{1101.5036}.

\bibtype{Article}%
\bibitem[{Riess} et al.(1998)]{Riess+1998}
\bibinfo{author}{{Riess} AG}, \bibinfo{author}{{Filippenko} AV}, \bibinfo{author}{{Challis} P}, \bibinfo{author}{{Clocchiatti} A}, \bibinfo{author}{{Diercks} A}, \bibinfo{author}{{Garnavich} PM}, \bibinfo{author}{{Gilliland} RL}, \bibinfo{author}{{Hogan} CJ}, \bibinfo{author}{{Jha} S}, \bibinfo{author}{{Kirshner} RP}, \bibinfo{author}{{Leibundgut} B}, \bibinfo{author}{{Phillips} MM}, \bibinfo{author}{{Reiss} D}, \bibinfo{author}{{Schmidt} BP}, \bibinfo{author}{{Schommer} RA}, \bibinfo{author}{{Smith} RC}, \bibinfo{author}{{Spyromilio} J}, \bibinfo{author}{{Stubbs} C}, \bibinfo{author}{{Suntzeff} NB} and  \bibinfo{author}{{Tonry} J} (\bibinfo{year}{1998}), \bibinfo{month}{Sep.}
\bibinfo{title}{{Observational Evidence from Supernovae for an Accelerating Universe and a Cosmological Constant}}.
\bibinfo{journal}{{\em \aj}} \bibinfo{volume}{116} (\bibinfo{number}{3}): \bibinfo{pages}{1009--1038}. \bibinfo{doi}{\doi{10.1086/300499}}.
\eprint{astro-ph/9805201}.

\bibtype{Article}%
\bibitem[{Rivinius} et al.(2013)]{Rivinius+2013}
\bibinfo{author}{{Rivinius} T}, \bibinfo{author}{{Carciofi} AC} and  \bibinfo{author}{{Martayan} C} (\bibinfo{year}{2013}), \bibinfo{month}{Oct.}
\bibinfo{title}{{Classical Be stars. Rapidly rotating B stars with viscous Keplerian decretion disks}}.
\bibinfo{journal}{{\em \aapr}} \bibinfo{volume}{21}, \bibinfo{eid}{69}. \bibinfo{doi}{\doi{10.1007/s00159-013-0069-0}}.
\eprint{1310.3962}.

\bibtype{Article}%
\bibitem[{Rocha} et al.(2024)]{Rocha+2024}
\bibinfo{author}{{Rocha} KA}, \bibinfo{author}{{Kalogera} V}, \bibinfo{author}{{Doctor} Z}, \bibinfo{author}{{Andrews} JJ}, \bibinfo{author}{{Sun} M}, \bibinfo{author}{{Gossage} S}, \bibinfo{author}{{Bavera} SS}, \bibinfo{author}{{Fragos} T}, \bibinfo{author}{{Kovlakas} K}, \bibinfo{author}{{Kruckow} MU}, \bibinfo{author}{{Misra} D}, \bibinfo{author}{{Srivastava} PM}, \bibinfo{author}{{Xing} Z} and  \bibinfo{author}{{Zapartas} E} (\bibinfo{year}{2024}), \bibinfo{month}{Aug.}
\bibinfo{title}{{To Be or Not To Be: The Role of Rotation in Modeling Galactic Be X-Ray Binaries}}.
\bibinfo{journal}{{\em \apj}} \bibinfo{volume}{971} (\bibinfo{number}{2}), \bibinfo{eid}{133}. \bibinfo{doi}{\doi{10.3847/1538-4357/ad5955}}.
\eprint{2403.07172}.

\bibtype{Article}%
\bibitem[{R{\"o}pke} and {De Marco}(2023)]{RopkeDeMarco2023}
\bibinfo{author}{{R{\"o}pke} FK} and  \bibinfo{author}{{De Marco} O} (\bibinfo{year}{2023}), \bibinfo{month}{Dec.}
\bibinfo{title}{{Simulations of common-envelope evolution in binary stellar systems: physical models and numerical techniques}}.
\bibinfo{journal}{{\em Living Reviews in Computational Astrophysics}} \bibinfo{volume}{9} (\bibinfo{number}{1}), \bibinfo{eid}{2}. \bibinfo{doi}{\doi{10.1007/s41115-023-00017-x}}.
\eprint{2212.07308}.

\bibtype{Inproceedings}%
\bibitem[{Sana}(2017)]{Sana2017}
\bibinfo{author}{{Sana} H} (\bibinfo{year}{2017}), \bibinfo{month}{Nov.}, \bibinfo{title}{{The multiplicity of massive stars: a 2016 view}}, \bibinfo{editor}{{Eldridge} JJ}, \bibinfo{editor}{{Bray} JC}, \bibinfo{editor}{{McClelland} LAS} and  \bibinfo{editor}{{Xiao} L}, (Eds.), \bibinfo{booktitle}{The Lives and Death-Throes of Massive Stars}, \bibinfo{series}{IAU Symposium}, \bibinfo{volume}{329},  \bibinfo{pages}{110--117}, \eprint{1703.01608}.

\bibtype{Article}%
\bibitem[{Sana} et al.(2012)]{Sana+2012}
\bibinfo{author}{{Sana} H}, \bibinfo{author}{{de Mink} SE}, \bibinfo{author}{{de Koter} A}, \bibinfo{author}{{Langer} N}, \bibinfo{author}{{Evans} CJ}, \bibinfo{author}{{Gieles} M}, \bibinfo{author}{{Gosset} E}, \bibinfo{author}{{Izzard} RG}, \bibinfo{author}{{Le Bouquin} JB} and  \bibinfo{author}{{Schneider} FRN} (\bibinfo{year}{2012}), \bibinfo{month}{Jul.}
\bibinfo{title}{{Binary Interaction Dominates the Evolution of Massive Stars}}.
\bibinfo{journal}{{\em Science}} \bibinfo{volume}{337} (\bibinfo{number}{6093}): \bibinfo{pages}{444}. \bibinfo{doi}{\doi{10.1126/science.1223344}}.
\eprint{1207.6397}.

\bibtype{Article}%
\bibitem[{Schneider} et al.(2019)]{Schneider+2019}
\bibinfo{author}{{Schneider} FRN}, \bibinfo{author}{{Ohlmann} ST}, \bibinfo{author}{{Podsiadlowski} P}, \bibinfo{author}{{R{\"o}pke} FK}, \bibinfo{author}{{Balbus} SA}, \bibinfo{author}{{Pakmor} R} and  \bibinfo{author}{{Springel} V} (\bibinfo{year}{2019}), \bibinfo{month}{Oct.}
\bibinfo{title}{{Stellar mergers as the origin of magnetic massive stars}}.
\bibinfo{journal}{{\em \nat}} \bibinfo{volume}{574} (\bibinfo{number}{7777}): \bibinfo{pages}{211--214}. \bibinfo{doi}{\doi{10.1038/s41586-019-1621-5}}.
\eprint{1910.14058}.

\bibtype{Article}%
\bibitem[{Schootemeijer} et al.(2018)]{Schootemeijer+2018}
\bibinfo{author}{{Schootemeijer} A}, \bibinfo{author}{{G{\"o}tberg} Y}, \bibinfo{author}{{de Mink} SE}, \bibinfo{author}{{Gies} D} and  \bibinfo{author}{{Zapartas} E} (\bibinfo{year}{2018}), \bibinfo{month}{Jul.}
\bibinfo{title}{{Clues about the scarcity of stripped-envelope stars from the evolutionary state of the sdO+Be binary system {\ensuremath{\varphi}} Persei}}.
\bibinfo{journal}{{\em \aap}} \bibinfo{volume}{615}, \bibinfo{eid}{A30}. \bibinfo{doi}{\doi{10.1051/0004-6361/201731194}}.
\eprint{1803.02379}.

\bibtype{Article}%
\bibitem[{Sen} et al.(2022)]{Sen+2022}
\bibinfo{author}{{Sen} K}, \bibinfo{author}{{Langer} N}, \bibinfo{author}{{Marchant} P}, \bibinfo{author}{{Menon} A}, \bibinfo{author}{{de Mink} SE}, \bibinfo{author}{{Schootemeijer} A}, \bibinfo{author}{{Sch{\"u}rmann} C}, \bibinfo{author}{{Mahy} L}, \bibinfo{author}{{Hastings} B}, \bibinfo{author}{{Nathaniel} K}, \bibinfo{author}{{Sana} H}, \bibinfo{author}{{Wang} C} and  \bibinfo{author}{{Xu} XT} (\bibinfo{year}{2022}), \bibinfo{month}{Mar.}
\bibinfo{title}{{Detailed models of interacting short-period massive binary stars}}.
\bibinfo{journal}{{\em \aap}} \bibinfo{volume}{659}, \bibinfo{eid}{A98}. \bibinfo{doi}{\doi{10.1051/0004-6361/202142574}}.
\eprint{2111.03329}.

\bibtype{Article}%
\bibitem[{Shenar} et al.(2023)]{Shenar+2023}
\bibinfo{author}{{Shenar} T}, \bibinfo{author}{{Wade} GA}, \bibinfo{author}{{Marchant} P}, \bibinfo{author}{{Bagnulo} S}, \bibinfo{author}{{Bodensteiner} J}, \bibinfo{author}{{Bowman} DM}, \bibinfo{author}{{Gilkis} A}, \bibinfo{author}{{Langer} N}, \bibinfo{author}{{Nicolas-Chen{\'e}} A}, \bibinfo{author}{{Oskinova} L}, \bibinfo{author}{{Van Reeth} T}, \bibinfo{author}{{Sana} H}, \bibinfo{author}{{St-Louis} N}, \bibinfo{author}{{de Oliveira} AS}, \bibinfo{author}{{Todt} H} and  \bibinfo{author}{{Toonen} S} (\bibinfo{year}{2023}), \bibinfo{month}{Aug.}
\bibinfo{title}{{A massive helium star with a sufficiently strong magnetic field to form a magnetar}}.
\bibinfo{journal}{{\em Science}} \bibinfo{volume}{381} (\bibinfo{number}{6659}): \bibinfo{pages}{761--765}. \bibinfo{doi}{\doi{10.1126/science.ade3293}}.
\eprint{2308.08591}.

\bibtype{Article}%
\bibitem[{Shenar} et al.(2024)]{Shenar+2024}
\bibinfo{author}{{Shenar} T}, \bibinfo{author}{{Bodensteiner} J}, \bibinfo{author}{{Sana} H} and  \bibinfo{author}{et. al.} (\bibinfo{year}{2024}), \bibinfo{month}{Oct.}
\bibinfo{title}{{Binarity at LOw Metallicity (BLOeM): A spectroscopic VLT monitoring survey of massive stars in the SMC}}.
\bibinfo{journal}{{\em \aap}} \bibinfo{volume}{690}, \bibinfo{eid}{A289}. \bibinfo{doi}{\doi{10.1051/0004-6361/202451586}}.
\eprint{2407.14593}.

\bibtype{Article}%
\bibitem[{Shiber} et al.(2024)]{Shiber+2024}
\bibinfo{author}{{Shiber} S}, \bibinfo{author}{{Chatzopoulos} E}, \bibinfo{author}{{Munson} B} and  \bibinfo{author}{{Frank} J} (\bibinfo{year}{2024}), \bibinfo{month}{Feb.}
\bibinfo{title}{{Betelgeuse as a Merger of a Massive Star with a Companion}}.
\bibinfo{journal}{{\em \apj}} \bibinfo{volume}{962} (\bibinfo{number}{2}), \bibinfo{eid}{168}. \bibinfo{doi}{\doi{10.3847/1538-4357/ad0e0a}}.
\eprint{2310.14603}.

\bibtype{Article}%
\bibitem[{Soberman} et al.(1997)]{Soberman+1997}
\bibinfo{author}{{Soberman} GE}, \bibinfo{author}{{Phinney} ES} and  \bibinfo{author}{{van den Heuvel} EPJ} (\bibinfo{year}{1997}), \bibinfo{month}{Nov.}
\bibinfo{title}{{Stability criteria for mass transfer in binary stellar evolution.}}
\bibinfo{journal}{{\em \aap}} \bibinfo{volume}{327}: \bibinfo{pages}{620--635}. \bibinfo{doi}{\doi{10.48550/arXiv.astro-ph/9703016}}.
\eprint{astro-ph/9703016}.

\bibtype{Article}%
\bibitem[{Soszy{\'n}ski} et al.(2016)]{Soszynksi+2016}
\bibinfo{author}{{Soszy{\'n}ski} I}, \bibinfo{author}{{Pawlak} M}, \bibinfo{author}{{Pietrukowicz} P}, \bibinfo{author}{{Udalski} A}, \bibinfo{author}{{Szyma{\'n}ski} MK}, \bibinfo{author}{{Wyrzykowski} {\L}}, \bibinfo{author}{{Ulaczyk} K}, \bibinfo{author}{{Poleski} R}, \bibinfo{author}{{Koz{\l}owski} S}, \bibinfo{author}{{Skowron} DM}, \bibinfo{author}{{Skowron} J}, \bibinfo{author}{{Mr{\'o}z} P} and  \bibinfo{author}{{Hamanowicz} A} (\bibinfo{year}{2016}), \bibinfo{month}{Dec.}
\bibinfo{title}{{The OGLE Collection of Variable Stars. Over 450 000 Eclipsing and Ellipsoidal Binary Systems Toward the Galactic Bulge}}.
\bibinfo{journal}{{\em \actaa}} \bibinfo{volume}{66} (\bibinfo{number}{4}): \bibinfo{pages}{405--420}. \bibinfo{doi}{\doi{10.48550/arXiv.1701.03105}}.
\eprint{1701.03105}.

\bibtype{Article}%
\bibitem[{Sravan} et al.(2019)]{Sravan+2019}
\bibinfo{author}{{Sravan} N}, \bibinfo{author}{{Marchant} P} and  \bibinfo{author}{{Kalogera} V} (\bibinfo{year}{2019}), \bibinfo{month}{Nov.}
\bibinfo{title}{{Progenitors of Type IIb Supernovae. I. Evolutionary Pathways and Rates}}.
\bibinfo{journal}{{\em \apj}} \bibinfo{volume}{885} (\bibinfo{number}{2}), \bibinfo{eid}{130}. \bibinfo{doi}{\doi{10.3847/1538-4357/ab4ad7}}.
\eprint{1808.07580}.

\bibtype{Article}%
\bibitem[{Stanway} et al.(2016)]{Stanway+2016}
\bibinfo{author}{{Stanway} ER}, \bibinfo{author}{{Eldridge} JJ} and  \bibinfo{author}{{Becker} GD} (\bibinfo{year}{2016}), \bibinfo{month}{Feb.}
\bibinfo{title}{{Stellar population effects on the inferred photon density at reionization}}.
\bibinfo{journal}{{\em \mnras}} \bibinfo{volume}{456} (\bibinfo{number}{1}): \bibinfo{pages}{485--499}. \bibinfo{doi}{\doi{10.1093/mnras/stv2661}}.
\eprint{1511.03268}.

\bibtype{Article}%
\bibitem[{Sun} et al.(2023)]{Sun+2023}
\bibinfo{author}{{Sun} M}, \bibinfo{author}{{Townsend} RHD} and  \bibinfo{author}{{Guo} Z} (\bibinfo{year}{2023}), \bibinfo{month}{Mar.}
\bibinfo{title}{{gyre\_tides: Modeling Binary Tides within the GYRE Stellar Oscillation Code}}.
\bibinfo{journal}{{\em \apj}} \bibinfo{volume}{945} (\bibinfo{number}{1}), \bibinfo{eid}{43}. \bibinfo{doi}{\doi{10.3847/1538-4357/acb33a}}.
\eprint{2301.06599}.

\bibtype{Article}%
\bibitem[{Tauris} and {Savonije}(1999)]{TaurisSavonije1999}
\bibinfo{author}{{Tauris} TM} and  \bibinfo{author}{{Savonije} GJ} (\bibinfo{year}{1999}), \bibinfo{month}{Oct.}
\bibinfo{title}{{Formation of millisecond pulsars. I. Evolution of low-mass X-ray binaries with P\_orb> 2 days}}.
\bibinfo{journal}{{\em \aap}} \bibinfo{volume}{350}: \bibinfo{pages}{928--944}. \bibinfo{doi}{\doi{10.48550/arXiv.astro-ph/9909147}}.
\eprint{astro-ph/9909147}.

\bibtype{Book}%
\bibitem[{Tauris} and {van den Heuvel}(2023)]{Tauris2023}
\bibinfo{author}{{Tauris} TM} and  \bibinfo{author}{{van den Heuvel} EPJ} (\bibinfo{year}{2023}).
\bibinfo{title}{{Physics of Binary Star Evolution. From Stars to X-ray Binaries and Gravitational Wave Sources}}.
\bibinfo{doi}{\doi{10.48550/arXiv.2305.09388}}.

\bibtype{Article}%
\bibitem[{Thompson} et al.(2012)]{Thompson2012}
\bibinfo{author}{{Thompson} SE}, \bibinfo{author}{{Everett} M}, \bibinfo{author}{{Mullally} F}, \bibinfo{author}{{Barclay} T}, \bibinfo{author}{{Howell} SB}, \bibinfo{author}{{Still} M}, \bibinfo{author}{{Rowe} J}, \bibinfo{author}{{Christiansen} JL}, \bibinfo{author}{{Kurtz} DW}, \bibinfo{author}{{Hambleton} K}, \bibinfo{author}{{Twicken} JD}, \bibinfo{author}{{Ibrahim} KA} and  \bibinfo{author}{{Clarke} BD} (\bibinfo{year}{2012}), \bibinfo{month}{Jul.}
\bibinfo{title}{{A Class of Eccentric Binaries with Dynamic Tidal Distortions Discovered with Kepler}}.
\bibinfo{journal}{{\em \apj}} \bibinfo{volume}{753} (\bibinfo{number}{1}), \bibinfo{eid}{86}. \bibinfo{doi}{\doi{10.1088/0004-637X/753/1/86}}.
\eprint{1203.6115}.

\bibtype{Article}%
\bibitem[{Thorne} and {Zytkow}(1975)]{ThorneZytkow1975}
\bibinfo{author}{{Thorne} KS} and  \bibinfo{author}{{Zytkow} AN} (\bibinfo{year}{1975}), \bibinfo{month}{Jul.}
\bibinfo{title}{{Red giants and supergiants with degenerate neutron cores.}}
\bibinfo{journal}{{\em \apjl}} \bibinfo{volume}{199}: \bibinfo{pages}{L19--L24}. \bibinfo{doi}{\doi{10.1086/181839}}.

\bibtype{Article}%
\bibitem[{Tout} et al.(2008)]{Tout+2008}
\bibinfo{author}{{Tout} CA}, \bibinfo{author}{{Wickramasinghe} DT}, \bibinfo{author}{{Liebert} J}, \bibinfo{author}{{Ferrario} L} and  \bibinfo{author}{{Pringle} JE} (\bibinfo{year}{2008}), \bibinfo{month}{Jun.}
\bibinfo{title}{{Binary star origin of high field magnetic white dwarfs}}.
\bibinfo{journal}{{\em \mnras}} \bibinfo{volume}{387} (\bibinfo{number}{2}): \bibinfo{pages}{897--901}. \bibinfo{doi}{\doi{10.1111/j.1365-2966.2008.13291.x}}.
\eprint{0805.0115}.

\bibtype{Article}%
\bibitem[{Tylenda} et al.(2011)]{Tylenda+2011}
\bibinfo{author}{{Tylenda} R}, \bibinfo{author}{{Hajduk} M}, \bibinfo{author}{{Kami{\'n}ski} T}, \bibinfo{author}{{Udalski} A}, \bibinfo{author}{{Soszy{\'n}ski} I}, \bibinfo{author}{{Szyma{\'n}ski} MK}, \bibinfo{author}{{Kubiak} M}, \bibinfo{author}{{Pietrzy{\'n}ski} G}, \bibinfo{author}{{Poleski} R}, \bibinfo{author}{{Wyrzykowski} {\L}} and  \bibinfo{author}{{Ulaczyk} K} (\bibinfo{year}{2011}), \bibinfo{month}{Apr.}
\bibinfo{title}{{V1309 Scorpii: merger of a contact binary}}.
\bibinfo{journal}{{\em \aap}} \bibinfo{volume}{528}, \bibinfo{eid}{A114}. \bibinfo{doi}{\doi{10.1051/0004-6361/201016221}}.
\eprint{1012.0163}.

\bibtype{Article}%
\bibitem[{Van} and {Ivanova}(2021)]{VanIvanova2021}
\bibinfo{author}{{Van} KX} and  \bibinfo{author}{{Ivanova} N} (\bibinfo{year}{2021}), \bibinfo{month}{Dec.}
\bibinfo{title}{{Constraining Progenitors of Observed Low-mass X-ray Binaries Using Convection and Rotation-Boosted Magnetic Braking}}.
\bibinfo{journal}{{\em \apj}} \bibinfo{volume}{922} (\bibinfo{number}{2}), \bibinfo{eid}{174}. \bibinfo{doi}{\doi{10.3847/1538-4357/ac236c}}.
\eprint{2109.06814}.

\bibtype{Article}%
\bibitem[{van den Heuvel} et al.(2017)]{vandenheuvel+2017}
\bibinfo{author}{{van den Heuvel} EPJ}, \bibinfo{author}{{Portegies Zwart} SF} and  \bibinfo{author}{{de Mink} SE} (\bibinfo{year}{2017}), \bibinfo{month}{Nov.}
\bibinfo{title}{{Forming short-period Wolf-Rayet X-ray binaries and double black holes through stable mass transfer}}.
\bibinfo{journal}{{\em \mnras}} \bibinfo{volume}{471} (\bibinfo{number}{4}): \bibinfo{pages}{4256--4264}. \bibinfo{doi}{\doi{10.1093/mnras/stx1430}}.
\eprint{1701.02355}.

\bibtype{Article}%
\bibitem[{Vanbeveren} et al.(1998)]{Vanbeveren1998}
\bibinfo{author}{{Vanbeveren} D}, \bibinfo{author}{{De Donder} E}, \bibinfo{author}{{Van Bever} J}, \bibinfo{author}{{Van Rensbergen} W} and  \bibinfo{author}{{De Loore} C} (\bibinfo{year}{1998}), \bibinfo{month}{Nov.}
\bibinfo{title}{{The WR and O-type star population predicted by massive star evolutionary synthesis}}.
\bibinfo{journal}{{\em \na}} \bibinfo{volume}{3} (\bibinfo{number}{7}): \bibinfo{pages}{443--492}. \bibinfo{doi}{\doi{10.1016/S1384-1076(98)00020-7}}.

\bibtype{Article}%
\bibitem[{Vinciguerra} et al.(2020)]{Vinciguerra+2020}
\bibinfo{author}{{Vinciguerra} S}, \bibinfo{author}{{Neijssel} CJ}, \bibinfo{author}{{Vigna-G{\'o}mez} A}, \bibinfo{author}{{Mandel} I}, \bibinfo{author}{{Podsiadlowski} P}, \bibinfo{author}{{Maccarone} TJ}, \bibinfo{author}{{Nicholl} M}, \bibinfo{author}{{Kingdon} S}, \bibinfo{author}{{Perry} A} and  \bibinfo{author}{{Salemi} F} (\bibinfo{year}{2020}), \bibinfo{month}{Nov.}
\bibinfo{title}{{Be X-ray binaries in the SMC as indicators of mass-transfer efficiency}}.
\bibinfo{journal}{{\em \mnras}} \bibinfo{volume}{498} (\bibinfo{number}{4}): \bibinfo{pages}{4705--4720}. \bibinfo{doi}{\doi{10.1093/mnras/staa2177}}.
\eprint{2003.00195}.

\bibtype{Article}%
\bibitem[{von Zeipel}(1924)]{vonZeipel1924}
\bibinfo{author}{{von Zeipel} H} (\bibinfo{year}{1924}), \bibinfo{month}{Jun.}
\bibinfo{title}{{The radiative equilibrium of a rotating system of gaseous masses}}.
\bibinfo{journal}{{\em \mnras}} \bibinfo{volume}{84}: \bibinfo{pages}{665--683}. \bibinfo{doi}{\doi{10.1093/mnras/84.9.665}}.

\bibtype{Article}%
\bibitem[{Webbink}(1984)]{Webbink1984}
\bibinfo{author}{{Webbink} RF} (\bibinfo{year}{1984}), \bibinfo{month}{Feb.}
\bibinfo{title}{{Double white dwarfs as progenitors of R Coronae Borealis stars and type I supernovae.}}
\bibinfo{journal}{{\em \apj}} \bibinfo{volume}{277}: \bibinfo{pages}{355--360}. \bibinfo{doi}{\doi{10.1086/161701}}.

\bibtype{Article}%
\bibitem[{Whelan} and {Iben}(1973)]{WhelanIben1973}
\bibinfo{author}{{Whelan} J} and  \bibinfo{author}{{Iben} Icko J} (\bibinfo{year}{1973}), \bibinfo{month}{Dec.}
\bibinfo{title}{{Binaries and Supernovae of Type I}}.
\bibinfo{journal}{{\em \apj}} \bibinfo{volume}{186}: \bibinfo{pages}{1007--1014}. \bibinfo{doi}{\doi{10.1086/152565}}.

\bibtype{Article}%
\bibitem[{Wilson} and {Devinney}(1971)]{WilsonDevinney1971}
\bibinfo{author}{{Wilson} RE} and  \bibinfo{author}{{Devinney} EJ} (\bibinfo{year}{1971}), \bibinfo{month}{Jun.}
\bibinfo{title}{{Realization of Accurate Close-Binary Light Curves: Application to MR Cygni}}.
\bibinfo{journal}{{\em \apj}} \bibinfo{volume}{166}: \bibinfo{pages}{605}. \bibinfo{doi}{\doi{10.1086/150986}}.

\bibtype{Article}%
\bibitem[{Yoon} et al.(2010)]{Yoon+2010}
\bibinfo{author}{{Yoon} SC}, \bibinfo{author}{{Woosley} SE} and  \bibinfo{author}{{Langer} N} (\bibinfo{year}{2010}), \bibinfo{month}{Dec.}
\bibinfo{title}{{Type Ib/c Supernovae in Binary Systems. I. Evolution and Properties of the Progenitor Stars}}.
\bibinfo{journal}{{\em \apj}} \bibinfo{volume}{725} (\bibinfo{number}{1}): \bibinfo{pages}{940--954}. \bibinfo{doi}{\doi{10.1088/0004-637X/725/1/940}}.
\eprint{1004.0843}.

\bibtype{Article}%
\bibitem[{Yoon} et al.(2017)]{Yoon+2017}
\bibinfo{author}{{Yoon} SC}, \bibinfo{author}{{Dessart} L} and  \bibinfo{author}{{Clocchiatti} A} (\bibinfo{year}{2017}), \bibinfo{month}{May}.
\bibinfo{title}{{Type Ib and IIb Supernova Progenitors in Interacting Binary Systems}}.
\bibinfo{journal}{{\em \apj}} \bibinfo{volume}{840} (\bibinfo{number}{1}), \bibinfo{eid}{10}. \bibinfo{doi}{\doi{10.3847/1538-4357/aa6afe}}.
\eprint{1701.02089}.

\bibtype{Article}%
\bibitem[{Zahn}(1975)]{Zahn1975}
\bibinfo{author}{{Zahn} JP} (\bibinfo{year}{1975}), \bibinfo{month}{Jul.}
\bibinfo{title}{{The dynamical tide in close binaries.}}
\bibinfo{journal}{{\em \aap}} \bibinfo{volume}{41}: \bibinfo{pages}{329--344}.

\bibtype{Article}%
\bibitem[{Zahn}(1977)]{Zahn1977}
\bibinfo{author}{{Zahn} JP} (\bibinfo{year}{1977}), \bibinfo{month}{May}.
\bibinfo{title}{{Tidal friction in close binary systems.}}
\bibinfo{journal}{{\em \aap}} \bibinfo{volume}{57}: \bibinfo{pages}{383--394}.

\bibtype{Article}%
\bibitem[{Zapartas} et al.(2017)]{Zapartas+2017}
\bibinfo{author}{{Zapartas} E}, \bibinfo{author}{{de Mink} SE}, \bibinfo{author}{{Izzard} RG}, \bibinfo{author}{{Yoon} SC}, \bibinfo{author}{{Badenes} C}, \bibinfo{author}{{G{\"o}tberg} Y}, \bibinfo{author}{{de Koter} A}, \bibinfo{author}{{Neijssel} CJ}, \bibinfo{author}{{Renzo} M}, \bibinfo{author}{{Schootemeijer} A} and  \bibinfo{author}{{Shrotriya} TS} (\bibinfo{year}{2017}), \bibinfo{month}{May}.
\bibinfo{title}{{Delay-time distribution of core-collapse supernovae with late events resulting from binary interaction}}.
\bibinfo{journal}{{\em \aap}} \bibinfo{volume}{601}, \bibinfo{eid}{A29}. \bibinfo{doi}{\doi{10.1051/0004-6361/201629685}}.
\eprint{1701.07032}.

\bibtype{Article}%
\bibitem[{Zhao} et al.(2008)]{Zhao+2008}
\bibinfo{author}{{Zhao} M}, \bibinfo{author}{{Gies} D}, \bibinfo{author}{{Monnier} JD}, \bibinfo{author}{{Thureau} N}, \bibinfo{author}{{Pedretti} E}, \bibinfo{author}{{Baron} F}, \bibinfo{author}{{Merand} A}, \bibinfo{author}{{ten Brummelaar} T}, \bibinfo{author}{{McAlister} H}, \bibinfo{author}{{Ridgway} ST}, \bibinfo{author}{{Turner} N}, \bibinfo{author}{{Sturmann} J}, \bibinfo{author}{{Sturmann} L}, \bibinfo{author}{{Farrington} C} and  \bibinfo{author}{{Goldfinger} PJ} (\bibinfo{year}{2008}), \bibinfo{month}{Sep.}
\bibinfo{title}{{First Resolved Images of the Eclipsing and Interacting Binary {\ensuremath{\beta}} Lyrae}}.
\bibinfo{journal}{{\em \apjl}} \bibinfo{volume}{684} (\bibinfo{number}{2}): \bibinfo{pages}{L95}. \bibinfo{doi}{\doi{10.1086/592146}}.
\eprint{0808.0932}.

\end{thebibliography*}

\end{document}